\documentclass[12pt]{article}

\usepackage[letterpaper, margin=1in]{geometry}
\usepackage{amsmath,mathtools, adjustbox}
\usepackage{booktabs}
\usepackage{subfigure}
\usepackage{kpfonts, xcolor, float}
\usepackage{natbib, tabularx}
\usepackage[hidelinks]{hyperref}
\usepackage[title]{appendix}
\usepackage{tkz-graph}  
\usepackage{tikz-cd}
\usetikzlibrary{shapes.geometric}%
\usepackage{makecell}
\usepackage{mathtools}
\usepackage{datatool}
\usepackage{caption}
\usepackage{placeins}
\usepackage{afterpage}
\usepackage{pdflscape}
\usepackage{bbm}

\usepackage[flushleft]{threeparttable}

\newtheorem{assumption}{Assumption}
\newtheorem{theorem}{Theorem}
\newtheorem{lemma}{Lemma}
\newtheorem{claim}{Claim}

\newcommand\fnote[1]{\captionsetup{font=small}\caption*{#1}}
\newcommand*{\getvar}[2][1]{%
    \DTLfetch{vartable}{name}{#2}{value}%
}
\DTLloaddb{vartable}{source/vartable.csv}
\makeatletter
  \def\title@font{\Large}
  \let\ltx@maketitle\@maketitle
  \def\@maketitle{\bgroup%
    \let\ltx@title\@title%
    \def\@title{\resizebox{\textwidth}{!}{%
      \mbox{\title@font\ltx@title}%
    }}%
    \ltx@maketitle%
  \egroup}
\makeatother

\title{Difference-in-Differences with Multiple Events}

\author{Lin-Tung Tsai\footnote{Stanford University. Many thanks to Kuan-Ming Chen and Natalia Serna for their comments.}}

\begin{document}
\begin{filecontents*}{source/vartable.csv}
name,value
bias_short,27.1
bias_avg,35.4
short_naive,-2.973
short_double,-1.087
post_naive,-3.333
post_double,1.858
\end{filecontents*}

\maketitle


\begin{abstract}

This paper studies staggered Difference-in-Differences (DiD) design when there is a second event confounding the target event. When the events are correlated, the treatment and the control group are unevenly exposed to the effects of the second event, causing an omitted event bias. To address this bias, I propose a two-stage DiD design. In the first stage, I estimate the combined effects of both treatments using a control group that is neither treated nor confounded. In the second stage, I isolate the effects of the target treatment by leveraging a parallel treatment effect assumption and a control group that is treated but not yet confounded. Finally, I apply this method to revisit the effect of minimum wage increases on teen employment using state-level hikes between 2010 and 2020. I find that the Medicaid expansion under the ACA is a significant confounder: controlling for this bias reduces the short-term estimate of the minimum wage effect by two-thirds.

\end{abstract}



In 1992, New Jersey not only increased its minimum wage from \$4.25 to \$5.05 per hour, but also reduced its sales tax from 7\% to 6\% \citep{King1992-mv}. So when \cite{Card1993-ej} measured a differential employment trend between New Jersey and Pennsylvania, was it from the minimum wage hike or the sales tax cut? When these confounding events correlates with the event of interest, its treatment effect confounds the trend of the treated group and the control group unevenly. This threatens the parallel trend assumption and creates an \textit{omitted event bias}.

\bigskip

This paper introduces a two-stage Difference-in-Differences estimator (Double DiD) that controls for confounding events while allowing for rich treatment effect heterogeneity. I introduce a second event to the group-time average treatment effect framework \citep[CS hereafter]{Callaway2021-vo}. I show that the omitted event bias comes from the problem of confounded control group and of confounded treated group. Th two problems are addressed with two DiD designs: The first DiD identify the combined effect of the two treatments with a control group not yet treated nor confounded; The second DiD isolates the treatment effect of the target event with a control group simultaneously treated but not yet confounded. I also provide aggregation, estimation, and inference procedures based on CS' results. The estimator is implemented in an open-source R package, \textit{fastdid.}\footnote{Available at CRAN and \url{https://github.com/TsaiLintung/fastdid}}

\bigskip

I apply Double DiD to revisit the effect of states' staggered raise of the minimum wage on teen employment \citep{Cengiz2019-vr, Callaway2021-vo}. From 2010 to 2020, states' minimum wage hikes are positively correlated with the Medicaid Expansion under the ACA, causing the unconditional CS DiD estimator to have a bias the size of \getvar{bias_avg}\% of the effect from Medicaid Expansion in the short term. This bias creates a spurious negative estimate: comparing the standard DiD with the Double DiD estimator, the estimated effect in the year after the minimum wage increase weakens from \getvar{short_naive} percentage points (p.p.) to \getvar{short_double} p.p. and loses statistical significance.

\bigskip

I contribute to the literature on confounding events by allowing for more treatment effect heterogeneity. To address the omitted event bias, 18\% of the papers in \textit{American Economic Review} using Two-way-fixed-effects (TWFE) control for confounding events   \citep{De_Chaisemartin2020-ke}. However, after the negative weight problem became known \citep{Goodman-Bacon2021-ze}, TWFE is now largely replaced by new estimators  \footnote{See \cite{Roth2022-cb} for a review of this sprawling literature.}. While the replacements are more robust to treatment effect heterogeneity, they are also less flexible in specification and cannot directly address confounding events. To the best of my knowledge, two existing methods in the modern DiD literature can account for confounding treatments: \cite{Caetano2024-dn} considers time-varying covariates and \cite{de-Chaisemartin2023-jb} extends \cite{De_Chaisemartin2020-ke} to incorporate multiple treatments. Double DiD complements these methods by being the most flexible with the treatment effect heterogeneity of both treatments. This is achieved through being the most selective with the control group, thus I trade some statistical power for a more credible parallel trend assumption. 

\bigskip

Additionally, \cite{Fricke2017-jo} and \cite{Roller2023-zy} both study a two-treatment DiD in a two-period setup, \cite{Hull2018-mv} study a two-period setup with multiple treatment arms. Finally, \cite{Matthay2022-tx} and \cite{Griffin2023-nu} document the correlation among state policies and their potential confoundedness. 
\section{Setup}

My setup extends the model of \cite{Callaway2021-vo} by adding a second event. Consider a panel with periods $t \in \{1,...,T\}$ and units $i \in \{1,...,N\} = \mathbb{I} $. $D^d_{it} \in \{0,1\}$ denotes whether unit $i$ is treated by treatment $d\in \{1,2\}$ in period $t$. Without loss of generality, I focus on identifying the effect from treatment 1 and call treatment 1 the target treatment and treatment 2 the confounding treatment. This two-treatment setup can also address multiple confounding treatments by setting $D^2_{it} = \max(D^2_{it},..., D^R_{it})$. 

\bigskip

Once a unit is treated, it will continue to be treated (treatment is an absorbing state): 

\begin{assumption}[Irreversible Treatments]\label{ass:irrtreat}
    For $d = 1,2$, $D^d_1 = 0$ almost surely (a.s.). For $d = 1,2, t = 2,..,T$ $D^d_{t-1} = 1 \implies D^d_{t} = 1$.
\end{assumption}

\noindent With irreversible treatments, the switch to being treated can be called an event. I call the period in which a unit experiences an event as its cohort, defined as the period when it was first treated by event $d$, $G^d_i = \underset{D^d_{it} = 1}{\min} t$. If a unit is never treated by event $d$, I set $G^d_i = \infty$. I also define an auxiliary indicator for whether the unit is in $g$ cohort, $G^d_{ig} = \mathbbm{1}\{G^d_i = g\}$. I denote the cohorts with some later-treated groups as $\mathbb{G}^d = \text{supp}(G^d)\setminus\{\bar{g}^d\}$ and the latest-treated among them as $\bar{g}^d = \max{G^d}$.

\bigskip

Let's set up the dynamic potential outcome \citep{Murphy2001-ha}. Let $Y_{it}(0,0) \in \mathbb{R}$ denote the unit $i$'s potential outcome if it is never treated by either treatment. Then, for $g^1, g^2 = 2,..., T$, $Y_{it}(g^1,0) \in \mathbb{R}$ is its potential outcome if first treated by event 1 in $g^1$ and never treated by event 2. $Y_{it}(0, g^2) \in \mathbb{R}$ is its potential outcome if first treated by event 2 in $g^2$ and never treated by event 1. Finally, $Y_{it}(g^1, g^2) \in \mathbb{R}$ is its potential outcome if first treated by event 1 in $g^1$ and first treated by event 2 in $G^1_i = g^1$. The realized potential outcome $Y_{it}$ is determined by which cohort the unit is actually in: 

\begin{equation}\label{eq:potout}
    Y_{it} = Y_{it}(0,0) + \sum^T_{g^1 = 2}\sum^T_{g^2 = 2}(Y_{it}(g^1, g^2)- Y_{it}(0, 0))G^1_{ig^1}G^2_{ig^2}.
\end{equation}

\noindent I assume that the observed outcome $Y_{it}$, treatments $D^1_{it}, D^2_{it}$, and covariates $X_i \in \mathbb{R}^{K}$ is randomly sampled across units: 

\begin{assumption}[Random Sampling]\label{ass:sample}
    $\{Y_{i1}, Y_{i2},...,Y_{iT}, X_i, D^1_{i1},D^1_{i2},...,D^1_{iT}, D^2_{i1},D^2_{i2},...,D^2_{iT}\}^N_{i=1}$ is independent and identically distributed (i.i.d.).
\end{assumption}

\noindent This imposes independent sampling across units but allows for arbitrary time-series autocorrelation. In the following parts, I suppress unit index $i$. 

\bigskip

My first substantial assumption is no anticipation of both treatments: 

\begin{assumption}[No Anticipation]\label{ass:anticipation}
    For $g^1\in \mathbb{G}^1, g^2\in \mathbb{G}^2, 1 \leq t < \max(g^1, g^2)$ 
    \begin{equation*}
        \mathbb{E}[Y_t(g^1, g^2)|X, G^1_{g^1} = 1, G^2_{g^2} = 1] = \begin{dcases*}
             \mathbb{E}[Y_t(0,0)|X, G^1_{g^1} = 1, G^2_{g^2} = 1], & if $ t < \min(g^1, g^2) $, \\
            \mathbb{E}[Y_t(g^1,0)|X, G^1_{g^1} = 1, G^2_{g^2} = 1], & if $g^1\leq t <g^2$, \\
            \mathbb{E}[Y_t(0,g^2)|X, G^1_{g^1} = 1, G^2_{g^2} = 1], & if $ g^2\leq t <g^1 $.
    \end{dcases*}  
    \end{equation*}

\end{assumption}

\noindent No anticipation says that the events can not have a treatment effect before the event occurs. This not only holds for the periods before either event ($t<g^1, t<g^2$) but also in periods after one event but before the other, e.g., $g^1\leq t < g^2$. I discuss the case of no anticipation only instead of allowing limited anticipation since it adds little insight and complicates the notation greatly. 

\bigskip

Next, the two-treatment version of the parallel trend assumption says that the potential outcome untreated by both events evolves parallelly between the treated units and units not treated by either treatment:

\begin{assumption}[Parallel Trends - Never-Treated]\label{ass:pt-never}
    For $g^1\in \mathbb{G}^1, g^2\in \mathbb{G}^2, t\in \{1,...,T\}$, and $t$ such that $t\geq \min(g^1, g^2)$

    $$ \mathbb{E}[Y_t(0, 0) - Y_{t-1}(0, 0)|X, G^1_{g^1} = 1, G^2_{g^2} = 1] = $$
    $$ \mathbb{E}[Y_t(0, 0) - Y_{t-1}(0, 0)|X, G^1_{\infty} = 1, G^2_{\infty} = 1].$$
\end{assumption}

\begin{assumption}[Parallel Trends - Not-Yet-Treated]\label{ass:pt}
    For $g^1\in \mathbb{G}^1, g^2\in \mathbb{G}^2, t\in \{1,...,T\}$, and $(s,t)$ such that $t\geq \min(g^1, g^2)$, $t \leq s < \min(\bar{G^1}, \bar{G^2})$

    $$ \mathbb{E}[Y_t(0, 0) - Y_{t-1}(0, 0)|X, G^1_{g^1} = 1, G^2_{g^2} = 1] = $$
    $$ \mathbb{E}[Y_t(0, 0) - Y_{t-1}(0, 0)|X, D^1_s = 0, D^2_s = 0, G^1_{g^1} = 0, G^2_{g^2} = 0].$$
\end{assumption}

\noindent Throughout this paper, I focus on the case of using not yet treated controls. Assumption \ref{ass:pt} is weaker than the one-treatment parallel trend assumption since it not only requires that the unit is not yet treated by the target treatment, but is also not yet treated by the confounding treatment. 

\bigskip

Finally, I impose the overlap condition: 

\begin{assumption}[Overlap]\label{ass:overlap}
    For $d\in \{1,2\}, t\in\{2,...,T\}, g^d\in \mathbb{G^d}$, exists $\epsilon>0$ such that $P(G^d_{g^d} = 1) > \epsilon$, $p_{g^1, t}(X, g^2) < 1-\epsilon$, and $p_{g^2, t}(X, g^1) < 1-\epsilon$, where $p_{g^1, t}(X, g^2) = P(G^1_{g^1} = 1|X, G^2 = g^2, G^1_{g^1}+(1-D^1_t)(1-G^1_{g^1} = 1))$ and $p_{g^2, t}(X, g^1) = P(G^2_{g^2} = 1|X, G^1 = g^1, G^2_{g^2}+(1-D^2_t)(1-G^2_{g^2} = 1)).$
\end{assumption}

\noindent This ensures that there is always some treated group and control group available and rules out irregular identification discussed in \cite{Khan2010-pu}. 

\FloatBarrier

\section{Identification}
\subsection{The Omitted Event Bias}

To understand how the DiD estimator is biased under confounding treatments, let's focus on the unconditional moments using not-yet-treated controls in CS (their equation 2.9): 

$$ATT^{CS}(g^1, t) \equiv \mathbb{E}[Y_t - Y_{g^1-1}|G^1_{g^1} = 1] - \mathbb{E}[Y_t - Y_{g^1-1}|D_t = 0],$$

\noindent which intends to identify the group-time average treatment effect of the target treatment:

$$ATT^1(g^1, t) \equiv \mathbb{E}[Y_t(g^1, 0) - Y_t(0,0)|G^1_{g^1} = 1].$$

\noindent With confounding events, $ATT^{CS}$ does not identify $ATT^1$. For exposition clarity, I shut down all treatment effect heterogeneity and denote the base period as $b = g^1-1$:

\begin{theorem}\label{thm:gtbias}
    Assuming $Y_t(g^1,0)-Y_t(0,0) = ATT^{1}$, $Y_t(0,g^2)-Y_t(0,0) = ATT^{2}$ and , $Y_t(g^1, g^2)-Y_t(0,0) = ATT$, for $g^1 \in \mathbb{G}^1, t\in \{2,...,T\}$ such that $g^1\leq t < \bar{g}^1$,
    
    \begin{align*}
    ATT^{CS}(g^1, t) =  &  ATT^{1} + \\
    & \underbrace{ATT^{2}\big\{P(D^2_t = 1, D^2_b = 0 | G^1_{g^1 }= 1) -P(D^2_t = 1, D^2_b = 0 | D^1_t = 0 )\big\} + }_{\text{omitted event bias}} \\
    & \underbrace{\{ATT-ATT^{1}-ATT^{2}\} P(D^2_t = 1 | G^1_{g^1} = 1).}_{\text{interaction bias}}
\end{align*}
\end{theorem}

\noindent All proofs are collected in Appendix \ref{app:proof}. Theorem \ref{thm:gtbias} shows two sources of bias. First, the omitted variable bias occurs because under confounding events, the treatment group and control group have non-parallel exposure to the treatment effect of the confounding treatment. Additionally, interaction bias occurs if the target treatment interacts with the confounding treatment ($ ATT^1 + ATT^2 \neq ATT$). This paper will focus on the omitted event bias and leave interaction bias for future research. 

\bigskip

\begin{figure}
\includegraphics[width=0.9\textwidth]{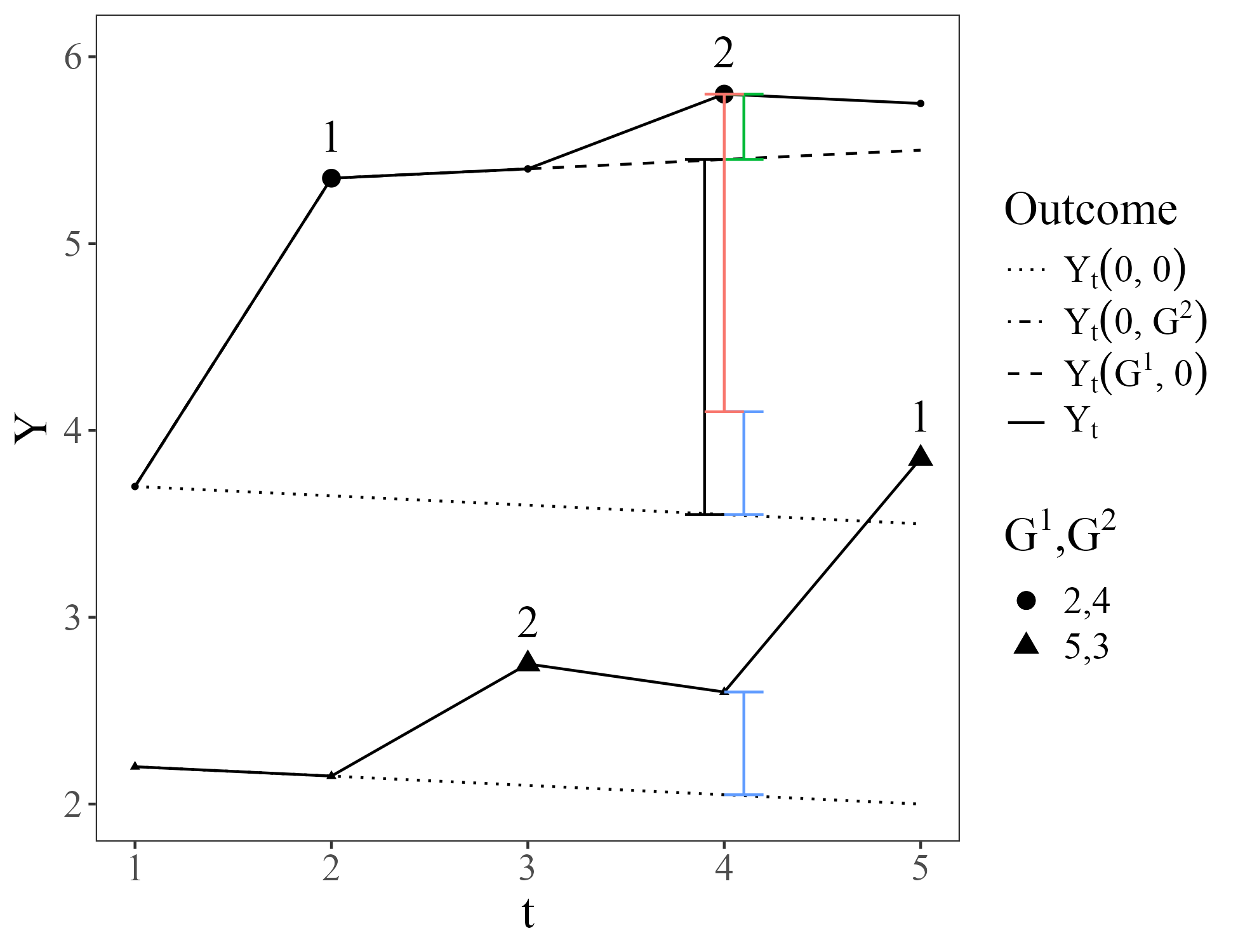} 
\caption{Omitted Event Bias in DiD}
    \label{fig:oeb}
    \fnote{This figure plots the outcome of two cohorts: a treated group that is treated in period 2 and confounded in period 4 ($(G^1, G^2) = (2,4)$), and a control group not yet treated by event 1 in period 2 $(G^1, G^2) = (5,3)$. A DiD estimator that uses period 1 as the base period (red bar) is biased from the target treatment effect (black bar) in two ways: the treated group is confounded by treatment 2 (green bar), and the control group is also confounded (blue bar).}
\end{figure}

Analogous to omitted variable bias, the size of the omitted event bias is determined by the size of the confounding treatment effect ($ATT^{2}$) and the correlation between two treatments:

\begin{equation}\label{eq:diag}
    \Gamma_{12}(g^1, t) \equiv P(D^2_t = 1, D^2_b = 0 | G^1_{g^1 }= 1) -P(D^2_t = 1, D^2_b = 0 | D^1_t = 0),
\end{equation}

\noindent which can be interpreted as the proportion such that $ATT^2$ contaminates the DiD estimates. Furthermore, the omitted event bias can be decomposed into two parts (see Figure \ref{fig:oeb} for an example):

\begin{equation*}
    \underbrace{ATT^{2}P(D^2_t = 1, D^2_b = 0 | G^1_{g^1 }= 1)}_{\text{confounded treatment}} -\underbrace{ATT^{2}P(D^2_t = 1, D^2_b = 0 | D^1_t = 0).}_{\text{confounded control}}
\end{equation*}

This shows two sources of bias. The control units ($D^2_t=0$) are confounded if the confounding event occurred between the base period and the target period ($D_t^2 = 1, D^2_b = 0 \iff t \geq G^2 > b$). Similarly, the treated units ($G^1_{g^1}=1$) can also be confounded. These two problems will dealt with using two DiD designs. 



\FloatBarrier
\subsection{DiD with Not Yet Treated Nor Confounded Controls}

The first stage DiD addresses the problem of confounded control units by using only units not yet treated by either treatment (not-yet-treated-nor-confounded) as controls. Let's first introduce the building block of the analysis, the \textit{double-group-time average treatment effect on the treated} (DGTATT), defined as the average treatment effect from both treatments among the units in the same event 1 cohort $G_{g^1}^1 = 1$, event 2 cohort $G_{g^2}^2 = 1$, and in period $t$:

\begin{equation}\label{eq:attggt}
    ATT(g^1, g^2, t) \equiv \mathbb{E}[Y_t(g^1, g^2) - Y_t(0,0)|G_{g^1}^1 = 1, G_{g^2}^2 = 1].
\end{equation}

\noindent I also define the target treatment effect, the parameter of interest of this paper:

\begin{equation}
    ATT^{1}(g^1, g^2, t) \equiv \mathbb{E}[Y_t(g^1, 0) - Y_t(0,0)|G_{g^1}^1 = 1, G_{g^2}^2 = 1],\\
\end{equation}

\noindent for the confounding treatment: 

\begin{equation}
ATT^{2}(g^1, g^2, t) \equiv \mathbb{E}[Y_t(0, g^2) - Y_t(0,0)|G_{g^1}^1 = 1, G_{g^2}^2 = 1].
\end{equation}

\noindent This first stage focuses on identifying the combined effect of both treatments $ATT(g^1, g^2, t)$. The discrepancy between $ATT$ and $ATT^1$ comes from confounded treated units and will be addressed in the second stage. 

\bigskip

The parallel trend assumption (Assumption \ref{ass:pt}) provides a straightforward choice for the control group: units neither-yet-treated by both events $D^1_t = 0, D^2_t = 0$. For example, with an unconditional parallel trend, we have

\begin{align*}
    ATT_{unc}(g^1, g^2, t) \equiv & \mathbb{E}[Y_t - Y_{b}|G^1_{g^1} = 1, G^2_{g^2} = 1]  - \mathbb{E}[Y_t - Y_{b}|D^1_t = 0, D^2_t = 0] \\
    = & \mathbb{E}[Y_t(g^1, g^2)|G^1_{g^1} = 1, G^2_{g^2} = 1]  - \mathbb{E}[Y_b(0, 0)|G^1_{g^1} = 1, G^2_{g^2} = 1] \\
    & - \mathbb{E}[Y_t(0, 0)- Y_{b}(0, 0)|D^1_t = 0, D^2_t = 0] && \text{Assumption \ref{ass:anticipation}} \\
    = & \mathbb{E}[Y_t(g^1, g^2)- Y_t(0,0)|G^1_{g^1} = 1, G^2_{g^2} = 1] \\
    & + \mathbb{E}[Y_t(0,0) - Y_b(0, 0)|G^1_{g^1} = 1, G^2_{g^2} = 1] \\
    & - \mathbb{E}[Y_t(0, 0)- Y_{b}(0, 0)|D^1_t = 0, D^2_t = 0] \\
    = & \mathbb{E}[Y_t(g^1, g^2)- Y_t(0,0)|G^1_{g^1} = 1, G^2_{g^2} = 1] && \text{Assumption \ref{ass:pt}} \\
    = & ATT(g^1, g^2, t)
\end{align*}

\noindent where $b = \min(g^1, g^2)-1$. 

\bigskip

My DGTATT connects closely with the group-time average treatment effect in CS since using units not-yet-treated-nor-confounded is synonymous to using units not-yet-treated by an treatment defined as treated-or-confounded. Specifically, given a target cohort $g^1, g^2$, define a subsample of the data $\mathbb{I}(g^1, g^2)$ by removing units that are first treated by either event in $g = \min(g^1, g^2)$ but not in the target cohort ($g^1, g^2$). Formally, let $\mathbb{I}(g^1, g^2) = \{i\in \mathbb{I}: W(g^1, g^2) = 1\}$, where

\begin{equation*}
W_i(g^1, g^2) = \begin{dcases*}
    1 & if $G^1_i = g^1, G^2_i = g^2$, \\
    0 & if $G^1_i \neq g^1, \min(G^1_i, G^2_i) = \min(g^1_i, g^2_i)$, \\
    0 & if $G^2_i \neq g^2, \min(G^1_i, G^2_i) = \min(g^1_i, g^2_i)$, \\
    1 & else.
\end{dcases*}
\end{equation*}

\noindent In this subsample, my assumptions imply CS assumptions:  

\begin{lemma}\label{lem:assequiv}

    Given the target cohort $g^1, g^2$, and assume $i\in \mathbb{I}(g^1, g^2)$, under $\delta = 0$, $g = \min(g^1, g^2)$ and

    \begin{equation*}
    Y_{it}(g) = \begin{dcases*}
        Y_{it}(G^1_i, G^2_i) & if $ g = \min(G^1_i, G^2_i) $, \\
        Y_{it}(g,g) & else,
    \end{dcases*}\end{equation*}
    
    \noindent CS's Assumptions 1, 2, 3, 6 holds. Furthermore, Assumption \ref{ass:pt-never} implies CS's Assumption 4, Assumption \ref{ass:pt} implies CS's Assumption 5. 

\end{lemma}

\noindent With Lemma \ref{lem:assequiv}, I can now introduce other identifying moments of the double-group-time average treatment effect by defining them as CS' corresponding moment in $\mathbb{I}(g^1, g^2)$, let

\begin{equation}\label{eq:ipw}
    ATT_{ipw}(g^1, g^2, t) \equiv\mathbb{E} \left[ \left( \frac{G^1_{g^1}G^2_{g^2}}{\mathbb{E}[G^1_{g^1}G^2_{g^2}]} -  \frac{ w^{comp}(g^1, g^2, t)}{\mathbb{E} [w^{comp}(g^1, g^2, t)]} \right)\left( Y_t - Y_{b} \right) \right],
\end{equation}

\begin{equation}\label{eq:or}
    ATT_{or}(g^1, g^2, t) \equiv\mathbb{E} \left[ \left( \frac{G^1_{g^1}G^2_{g^2}}{\mathbb{E}[G^1_{g^1}G^2_{g^2}]}\right)\left( Y_t - Y_{b} - m^{ny}_{g^1, g^2,t}(X) \right) \right],
\end{equation}

\begin{equation}\label{eq:dr}
    ATT_{dr}(g^1, g^2, t) \equiv\mathbb{E} \left[ \left( \frac{G^1_{g^1}G^2_{g^2}}{\mathbb{E}[G^1_{g^1}G^2_{g^2}]} -  \frac{ w^{comp}(g^1, g^2, t)}{\mathbb{E} [w^{comp}(g^1, g^2, t)]} \right)\left( Y_t - Y_{b} - m^{ny}_{g^1, g^2,t}(X) \right) \right],
\end{equation}

\noindent with the base period $b=\min(g^1, g^2) - 1$, the two-event version of the propensity score $p_{g^1, g^2, t}(X) \equiv P(G^1_{g^1} = 1, G^2_{g^2} = 1|X, G^1_{g^1}G^2_{g^2}+(1-D_t)(1-G^1_{g^1}G^2_{g^2}) = 1)$, inverse probability weights $w^{comp}(g^1, g^2, t) \equiv \frac{p_{g^1, g^2, t}(X)(1-D_{t})(1-G^1_{g^1}G^2_{g^2})}{1-p_{g^1, g^2,t}(X)}, $ and outcome's control function $m^{ny}_{g^1, g^2, t} \equiv \mathbb{E}[Y_t - Y_{b}|X, D_t = 0, G^1_{g^1} = 0, G^2_{g^2} = 0]$. Therefore, their values are the same in either the full sample $\mathbb{I}$ or the subsample $\mathbb{I}(g^1, g^2)$. 

\bigskip

By showing that these moments are equivalent to CS's corresponding moments in the subsample $\mathbb{I}(g^1, g^2)$, Theorem 1 in CS provides identification of the double-group-time average treatment effect:

\begin{theorem}\label{thm:ggtident}

For $g^1 \in \mathbb{G}^1, g^2 \in \mathbb{G}^2, t\in \{2,...,T\}$ such that $\min(g^1, g^2) \leq t < \min(\bar{g}^1, \bar{g}^2)$,
$$ATT(g^1, g^2, t) = ATT_{unc}(g^1, g^2, t) = ATT_{ipw}(g^1, g^2, t) = ATT_{or}(g^1, g^2, t) = ATT_{dr}(g^1, g^2, t).$$
\end{theorem}

\noindent Note that these moments use none of the units excluded from $\mathbb{I}(g^1, g^2)$, i.e., units first treated by either on or before $\min(g^1, g^2)$ but yet are not in cohort $(g^1, g^2)$. With the confounded controls addressed, the remaining problem is the confounded treated group.

\subsection{Double DiD and Parallel Treatment Effects}

The second stage DiD addresses the problem of confounded treated units. The combined treatment effect $ATT$ identified by the first DiD (Theorem \ref{thm:ggtident}) does not always equal the target $ATT^1$ because it switches depending on the order of the events. By Assumption \ref{ass:anticipation} (no anticipation), we have 

\begin{equation}\label{eq:attswitch}
    ATT(g^1, g^2, t) = \begin{dcases*}
    ATT^{1}(g^1, g^2, t) & if $t \geq g^1, t < g^2 $\\
    ATT^{2}(g^1, g^2, t) & if $t < g^1, t \geq g^2$ \\
    ATT(g^1, g^2, t) & if $t \geq g^1, t \geq g^2$ \\
    0 & else.
\end{dcases*}
\end{equation}

\noindent In periods after event 1 and before event 2, $t\geq g^1, t<g^2$, we directly observe the target parameter $ATT^1$. However, if the unit is also confounded ($t\geq g^2$), we observe instead the combined treatment effect $ATT$. Therefore, our goal is to impute these missing treatment effects with a new parallel trend assumption. I again focus on the `not-yet` version:

\begin{assumption}[Parallel Treatment Effects]\label{ass:ptet}

    For $g^1 \in \mathbb{G}^1, g^2 \in \mathbb{G}^2, t\in \{1,...,T\}$, and $(s^1,t)$ such that $t\geq g^2$, $t \leq s^1 $
    $$ATT^{2}(g^1, g^2, t) - ATT^{2}(g^1, g^2, g^1-1) = ATT^{2}(s^1, g^2, t) - ATT^{2}(s^1, g^2, g^1-1),$$ 
    and for $(s^2,t)$ such that $t\geq g^1$, $t \leq s^2$
    $$ATT^{1}(g^1, g^2, t) - ATT^{1}(g^1, g^2, g^2-1) = ATT^{1}(g^1, s^2, t) - ATT^{1}(g^1, s^2, g^2-1).$$
\end{assumption}

This assumption sets units simultaneously treated but not yet confounded as the control group. By only using the units in the same own-cohort, I'm allowing the target treatment effect ($ATT^1$) to vary freely across the treatment cohort ($g^1$), and assuming it to be parallel across the confounding cohort ($g^2$). For the confounding treatment effect its the converse: free across confounding cohort and parallel across treatment cohort.  

\bigskip

Assumption \ref{ass:ptet} plays a similar role to the common trend assumption (Assumption 4) in \cite{de-Chaisemartin2023-jb}. Their assumption restricts the outcome trend to be parallel conditional on the treatment status from the last period  ($D^1_{t-1}, D^2_{t-1}$). This means their control group is all units with the same treatment status in the last period whereas I restrict it to the units in the same own-cohort. Being more selective with control groups means that my estimator will be robust to treatment effect heterogeneity across cohorts. \cite{Caetano2024-dn} focus on covariates in DiD and doesn't explicitly discuss confounding events. Still, their method can control for confounding events by viewing them as a time-varying covariate. In such case, since their framework extends to CS', full cohort and time heterogeneity for the target treatment is allowed. Yet how the estimator behaves under treatment effect heteroegenity of the confounding event is not discussed. 

\bigskip

Finally, for additive imputation, I need to shut down the interaction between the treatment effects: 

\begin{assumption}[Additive Effects]\label{ass:addeff}
    $$ATT(g^1, g^2, t) = ATT^{1}(g^1, g^2, t) + ATT^{2}(g^1, g^2, t).$$
\end{assumption}

\begin{figure}
\includegraphics[width=0.9\textwidth]{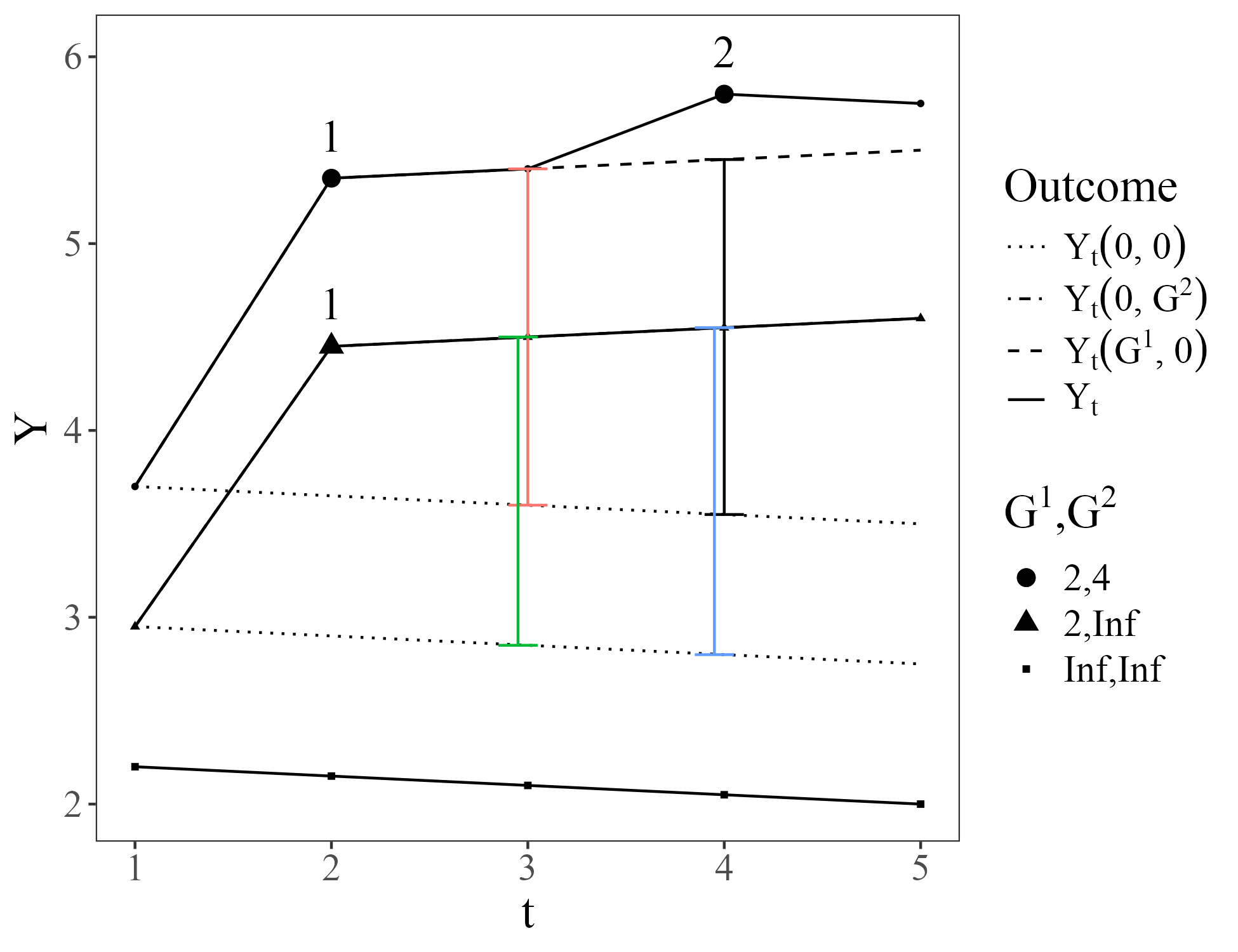} 
\caption{Imputation for the Treated-then-confounded}
    \label{fig:impute}
    \fnote{This figure plots the outcome of three cohorts: a treated group that experiences event 1 in period 2 and event 2 in period 4 ($(G^1, G^2) = (2,4)$), a control group for imputing the target treatment effect ($(G^1, G^2) = (2,\inf)$), and a control group for imputing the untreated potential outcome ($(G^1, G^2) = (\inf,\inf)$). After identifying the DGTATTs (colorful bars) with Theorem \ref{thm:ggtident}, the double DiD design uses the treatment effect trend of the control group (blue bar $-$ green bar) to impute the trend of the target treatment effect. Adding the trend and the level at the base period (red bar) gives the target treatment effect (black bar) under the parallel treatment effect assumption.}
\end{figure}

I have two strategies for identifying the target treatment effect depending on the order of the events. First, the DiD strategy (see Figure \ref{fig:impute} for example) is for units first treated then confounded ($g^1 > g^2$). I use the units in the same event 1 cohorts but not yet confounded by event 2 ($s^2>t$) as the control group and the period before event 2 ($b = g^2 -1$) as the base period. Then, a DiD on the treatment effects identified by Theorem \ref{thm:ggtident} can be used to identify the target treatment effect. Given a identifying moment $ATT_{id}$ such that $ATT = ATT_{id}$, let

\begin{equation}\label{eq:did}
    ATT^{did}_{id}(\tilde{g}^1, \tilde{g}^2, \tilde{t}) = \sum_{g^1\in \mathbb{G}^1}\sum_{g^2\in \mathbb{G}^2}\sum_{t=2}^Tw^{did}_{\tilde{g}^1, \tilde{g}^2, \tilde{t}}(g^1, g^2, t)ATT_{id}(g^1, g^2, t),
\end{equation}

\begin{equation*}
    w^{did}_{\tilde{g}^1, \tilde{g}^2, \tilde{t}}(g^1, g^2, t) \equiv \begin{dcases*}
    1, & if $t = \tilde{t}, g^1 = \tilde{g}^1, g^2 = \tilde{g}^2$, \\
    -1, & if $t = \tilde{g}^1-1, g^1 = \tilde{g}^1, g^2 = \tilde{g}^2$, \\
    -P(G^1_{g^1}=1 | G^2_{g^2} = 1, D^1_t = 0), & if $g^1 > t, g^2 = \tilde{g}^2, t = \tilde{t}$,\\
    P(G^1_{g^1}=1 | G^2_{g^2} = 1, D^1_t = 0), & if $g^1 > t, g^2 = \tilde{g}^2, t = \tilde{g}^1-1$,\\
    0 & else.
    \end{dcases*} 
\end{equation*}

\bigskip

\begin{figure}
 \includegraphics[width=0.9\textwidth]{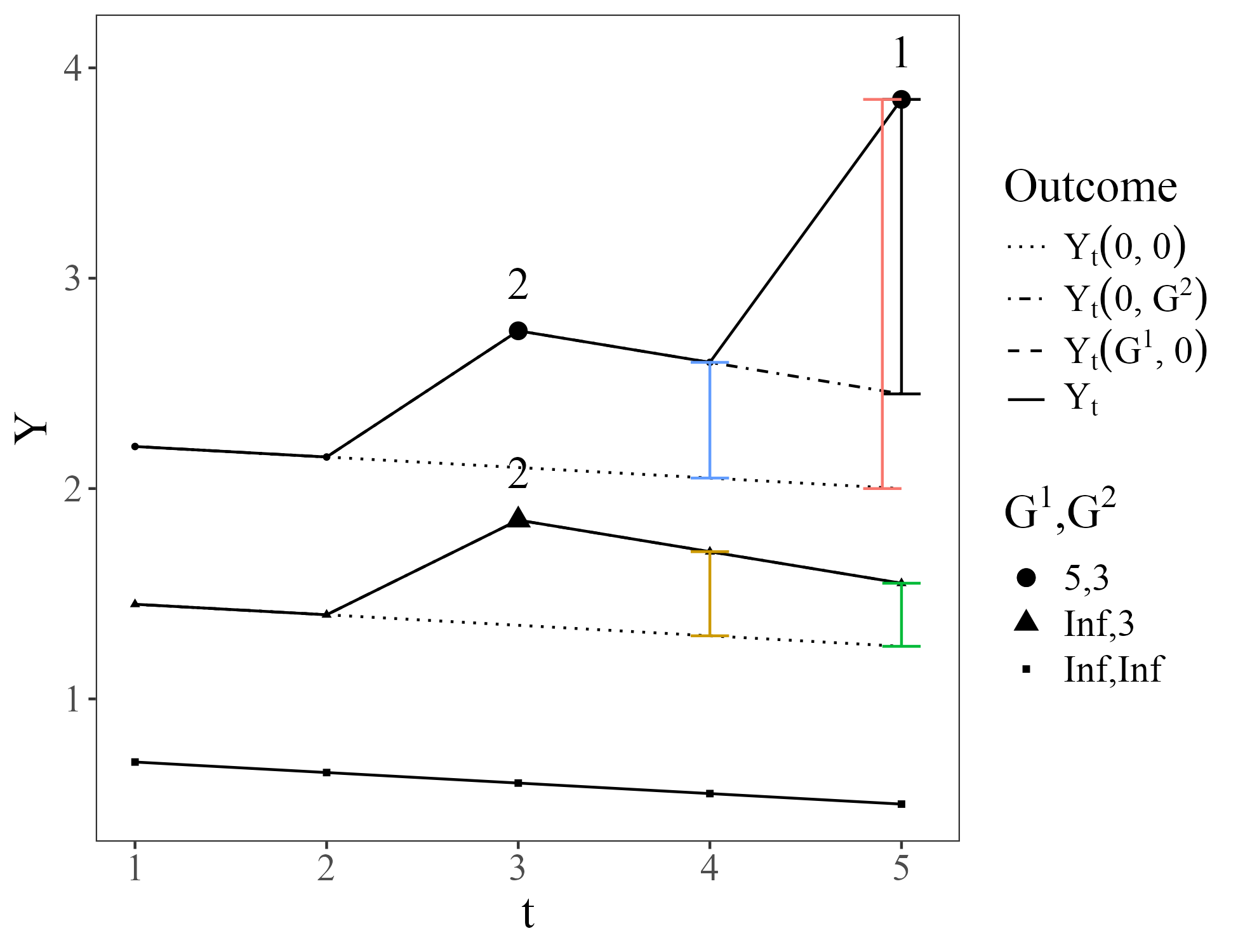} 
 \caption{Double DiD for the Confounded-then-treated}
    \label{fig:ddid}
    \fnote{This figure plots the outcome of three cohorts: a treated group that experiences event 1 in period 5 and event 2 in period 3 ($(G^1, G^2) = (5,3)$), a control group for imputing the confounding treatment effect ($(G^1, G^2) = (\inf,3)$), and a control group for imputing the untreated potential outcome ($(G^1, G^2) = (\inf,\inf)$). After identifying the DGTATTs (colorful bars) with Theorem \ref{thm:ggtident}, the double DiD design compares the treatment effect trend of the control group (green bar $-$ yellow bar) and the treated group (red bar $-$ blue bar). The difference is the target treatment effect (black bar) under the parallel treatment effect assumption. } 
\end{figure}

The second is imputation strategy (see Figure \ref{fig:ddid} for example) for units first confounded by event 2 and then treated by event 1 ($g^1 > g^2$). I use units in the same event 2 cohorts but not yet treated by event 1 ($s^1>t$) as the control group, and the period before event 1 ($b = g^1 -1$) as the base period. Then, the difference in the combined treatment effect trends between the control and treatment groups identifies the target treatment effect. 

\begin{equation}\label{eq:imp}
    ATT^{imp}_{id}(\tilde{g}^1, \tilde{g}^2, \tilde{t}) = \sum_{g^1\in \mathbb{G}^1}\sum_{g^2\in \mathbb{G}^2}\sum_{t=2}^Tw^{imp}_{\tilde{g}^1, \tilde{g}^2, \tilde{t}}(g^1, g^2, t)ATT_{id}(g^1, g^2, t),
\end{equation}

\begin{equation*}
    w^{imp}_{\tilde{g}^1, \tilde{g}^2, \tilde{t}}(g^1, g^2, t) \equiv \begin{dcases*}
    1, & if $t = \tilde{t}, g^1 = \tilde{g}^1, g^2 = \tilde{g}^2$, \\
    P(G^2_{g^2}=1 | G^1_{g^1} = 1, D^2_t = 0), & if $g^1 = \tilde{g}^1, g^2 > \tilde{t}, t = \tilde{t}$,\\
    -P(G^2_{g^2}=1 | G^1_{g^1} = 1, D^2_t = 0), & if $g^1 = \tilde{g}^1, g^2 > \tilde{t}, t = \tilde{g}^2-1$,\\
    0 & else.
    \end{dcases*} 
\end{equation*}

\bigskip

Combining these two strategies and the direct identification for periods before the confounding event, we can identify the target treatment effects $ATT^1$ for units with the discrepancy in treatment timings ($g^1 \neq g^2$) in periods with suitable control groups:

\begin{theorem}\label{thm:doubledid}
 Under assumption \ref{ass:ptet} and \ref{ass:addeff}, for an identifying moments $ATT_{id}$ such that $ATT_{id} = ATT$, for all $g^1, g^2, t$ such that (1) $g^1 < g^2$, $g^1 \leq t < \bar t^1$, or (2) $g^1 > g^2$, $g^1 \leq t < \bar t^2$, define the double DiD parameter $ATT^{dd}$ as:

\begin{equation}\label{eq:ddid}
    ATT^{dd}_{id}(g^1, g^2, t) \equiv \begin{dcases*}
    ATT_{id}(g^1, g^2, t), & if $ t < g^2 $, \\
    ATT^{imp}_{id}(g^1, g^2, t), & if $ t \geq g^2$ and $g^1 < g^2$,\\
    ATT^{did}_{id}(g^1, g^2, t), & else.
    \end{dcases*}  
\end{equation}

\noindent Then,

\begin{align*}
ATT^{1}(g^1, g^2, t) = ATT^{dd}_{id}(g^1, g^2, t),
\end{align*}

\noindent where the periods with control groups are $\bar t^1 \equiv \underset{G^1_{g^1} = 1}{\max} G^2$, $\bar t^2 \equiv \underset{G^2_{g^2} = 1}{\max} G^1.$

\end{theorem}

\noindent Similar to the pre-trend in a single-event setup, the treatment effect in periods before event 1 $t < g^1-1$ is overidentified due to the parallel trend assumption. Hence the estimator can be estimated at the pre-periods to diagnose potential violation of the parallel treatment effects assumption. One limitation is that for units treated and confounded in the same period ($g^1 = g^2$), we cannot identify the target treatment effect due to the lack of a base period such that we observe only one effect. 

\FloatBarrier

\section{Aggregation, Estimation, and Inference}

Given the connection in identification, the aggregation, estimation, and inference results are also similar to CS. To aggregate the double-group-time target treatment effects from the second stage, each summary parameter $\theta$ is associated with a weighting scheme, $w(g^1, g^2, t)$ such that

\begin{equation}\label{eq:agg}
    \theta = \sum_{g^1\in \mathbb{G}^1}\sum_{g^2\in \mathbb{G}^2}\sum_{t=2}^Tw(g^1, g^2, t)ATT^1(g^1, g^2, t).
\end{equation}

 \noindent For example, to obtain the group-time average treatment effect, $ATT^1(\tilde{g}^1, t)$, there is the weighting scheme 

\begin{equation}
   w_{\tilde g^1, t}(g^1, g^2, t) = \mathbbm{1}\{\tilde{g} = g\} \mathbbm{1}\{\tilde{t} = t\} A(g^1, g^2. t) P(G^2 = g^2| G^1 = \tilde{g}^1, A(g^1, g^2. t) =1), 
\end{equation}

\noindent where $A(g^1, g^2, t) \equiv \mathbbm{1}\{g^1<g^2\}\mathbbm{1}\{g^1 \leq t < \bar t^2\}+\mathbbm{1}\{g^1 > g^2\}\mathbbm{1}\{g^1 \leq t < \bar t^1\}$ is the indicator for availability of suitable control group. With the group-time parameter, the other aggregated parameters can be done using the summary scheme in CS. 

\bigskip

An additional aggregation scheme is the dynamic-staggered heterogeneity, i.e., how the treatment effect dynamics $e^1 = t-g^1$ differ by the timing difference of the two events $s^{12}$. For $\theta_{e^1, s^{12}} = \mathbb{E}[Y_t(g^1, 0)- Y_t(0,0)|t-g^1 = e^1, g^1 - g^2 = s^{12}]$, we can use the weighting scheme 

\begin{align}
\begin{split}
    w_{e^1, s^{12}}(g^1, g^2, t) = & \mathbbm{1}\{\tilde{e}^1 + g^1 \leq T \} \mathbbm{1}\{\tilde{e}^1 = t - g^1 \} \mathbbm{1}\{\tilde{s}^{12} = g^1 - g^2\} \\
    & P(G^1= g^1, G^2 = g^2| G^1+e^1 \leq T, G^1 - G^2 = \tilde{s}^{12}).
\end{split}
\end{align}

\bigskip

For estimation, the first stage moments from Theorem \ref{thm:ggtident} can be estimated with sample plug-ins. Following CS, I focus on the doubly-robust version since it has the benefit of additional robustness to misspecification and the arguments apply analogously to other parameters. For a generic $C \in \{0,1\}, Z$, let $\mathbb{E}_{n}[Z|C]$ denote the conditional summation $(\sum_{i=1}^NC_i)^{-1}\sum_{i=1}^NC_iZ_i$. Let

\begin{equation}\label{eq:drest}
    \widehat{ATT}_{dr}(g^1, g^2, t) \equiv \mathbb{E}_{n}[(\hat{w}^{treat}_{\min(g^1, g^2)}-\hat{w}^{comp}_{\min(g^1, g^2)}(Y_t-Y_{\min(g^1, g^2)-1}-\hat{m}_{\min(g^1, g^2), t}(X))|i\in \mathbb{I}(g^1, g^2)]
\end{equation}

where 

\begin{equation*}
    \hat{w}^{treat}_g = \frac{G_g}{\mathbb{E}_{n}[G_g|i\in \mathbb{I}(g^1, g^2)]}, \hat{w}_g^{comp, ny}=\frac{\frac{\hat{p}_{g,t}(X)(1-D_t)(1-G_g)}{1-\hat{p}_{g,t}(X)(1-D_t)(1-G_g)}}{\mathbb{E}_{n}\left[\frac{\hat{p}_{g,t}(X)(1-D_t)(1-G_g)}{1-\hat{p}_{g,t}(X)(1-D_t)(1-G_g)}|i\in \mathbb{I}(g^1, g^2)\right]},
\end{equation*}

\noindent and $\hat{p}(\cdot), \hat{m}(\cdot)$ is some estimators of ${p}(\cdot), (m)(\cdot)$.

\bigskip

For the second stage moments in Theorem \ref{thm:doubledid}, the formulation in Equation \ref{eq:did} and \ref{eq:imp} means that we can treat them as a summary parameter of the first stage estimates and aggregate them as weighted summations similar to Equation \ref{eq:agg}. 

\bigskip

Since the double-group-time average treatment effect estimators are equivalent to CS' estimators for group-time average treatment effects in $\mathbb{I}(g^1, g^2)$ (for example, Equation \ref{eq:drest} corresponds to Equation 4.2 in CS), and the double DiD estimators and other summary estimators are their aggregations CS' inference procedure in Section 4.1 applies. In summary, by assuming that the control function and propensity score can be estimated by smooth parametric models with $\sqrt{N}$ linear representation (CS Assumption 7) and some weak integrability conditions that ensure the second moments are finite (CS Assumption 8), CS Theorem 2 provides the influence function for calculating the standard error and its asymptotic normality. Then, CS Theorem 3 establishes the validity of the multiplier bootstrap for calculating the standard errors. The estimator is implemented in the R package \textit{fastdid}. 
\FloatBarrier
\section{Application: Minimum Wage and Teen Employment}

To illustrate the empirical relevance of omitted event bias, I revisit the effect of minimum wage on teen employment with the DiD design using staggered timing of minimum wage increases across the states in the U.S. \citep{Neumark2006-re, Dube2016-pr, Cengiz2019-vr, Callaway2021-vo}. My specification largely follows \cite{Callaway2021-vo}. I define treatment cohorts by the period when a state first increased its minimum wage and use the other states that do not increase the minimum wage as control groups (never-treated). Our specifications have two differences. First, to include the confounding event, Medicaid Expansion under the ACA, I use data from 2010 to 2020 instead of from 2001 to 2007. Similar to 2001 to 2007, the federal minimum wage is fixed at \$7.25 per hour during this period. Second, CS clustered their standard errors at the county level, whereas I clustered them at the state level, which is the level of the treatment as recommended in  \cite{Abadie2023-yl}.

\bigskip

I construct a state-year level dataset by linking several data sources, including state minimum wage history compiled by \cite{Sorens2008-ek}, the timing of Medicaid Expansion from \cite{KFF2024-cb}, teen employment from  Quarterly Workforce Indicators \citep{Abowd2009-ul}, and other state characteristics from \cite{University-of-Kentucky-Center-for-Poverty-Research2024-su}. I define yearly teen employment as the logarithm of the average count of employed teenagers (age 14-18) in the state across four quarters. Alaska, Mississippi, and Massachusetts are removed from the sample due to missing observations on teen employment from the QWI. Table \ref{tab:sumtable} reports the summary statistics of states by treatment. We can see that the states that increased minimum wage are also more likely to adopt Medicaid expansion by 2020, which may be explained by the different political leanings proxied by whether the governor is a Democrat. 

\begin{table}

\caption{\label{tab:sumtable}State Characteristics}
\centering
\begin{threeparttable}
\begin{tabular}[t]{lll}
\toprule
 & Control & Treated\\
\midrule
Poverty rate & 13.1 (3.644) & 12.4 (3.090)\\
Income per capita (k) & 45,208.0 (7,188.5) & 49,105.8 (9,173.6)\\
Medicaid expansion & 0.579 (0.495) & 0.964 (0.186)\\
Democratic governor & 0.206 (0.405) & 0.558 (0.497)\\
Population (k) & 5,824.2 (5,554.3) & 6,707.3 (7,743.9)\\
Count & 19 & 28\\
\bottomrule
\end{tabular}
\begin{tablenotes}
\item Note: Table reports summary statistics of states in the U.S. from 2010 to 2020 (except Alaska and Massachusetts). The sample mean is reported and the standard errors in parentheses. Treated are states that increased the minimum wage in the time period, and control are those that did not. Medicaid expansion is whether the state adopts Medicaid expansion under the ACA act in the periods.
\end{tablenotes}
\end{threeparttable}
\end{table}

\bigskip

\begin{figure}
    \centering
    \includegraphics[width=0.9\textwidth]{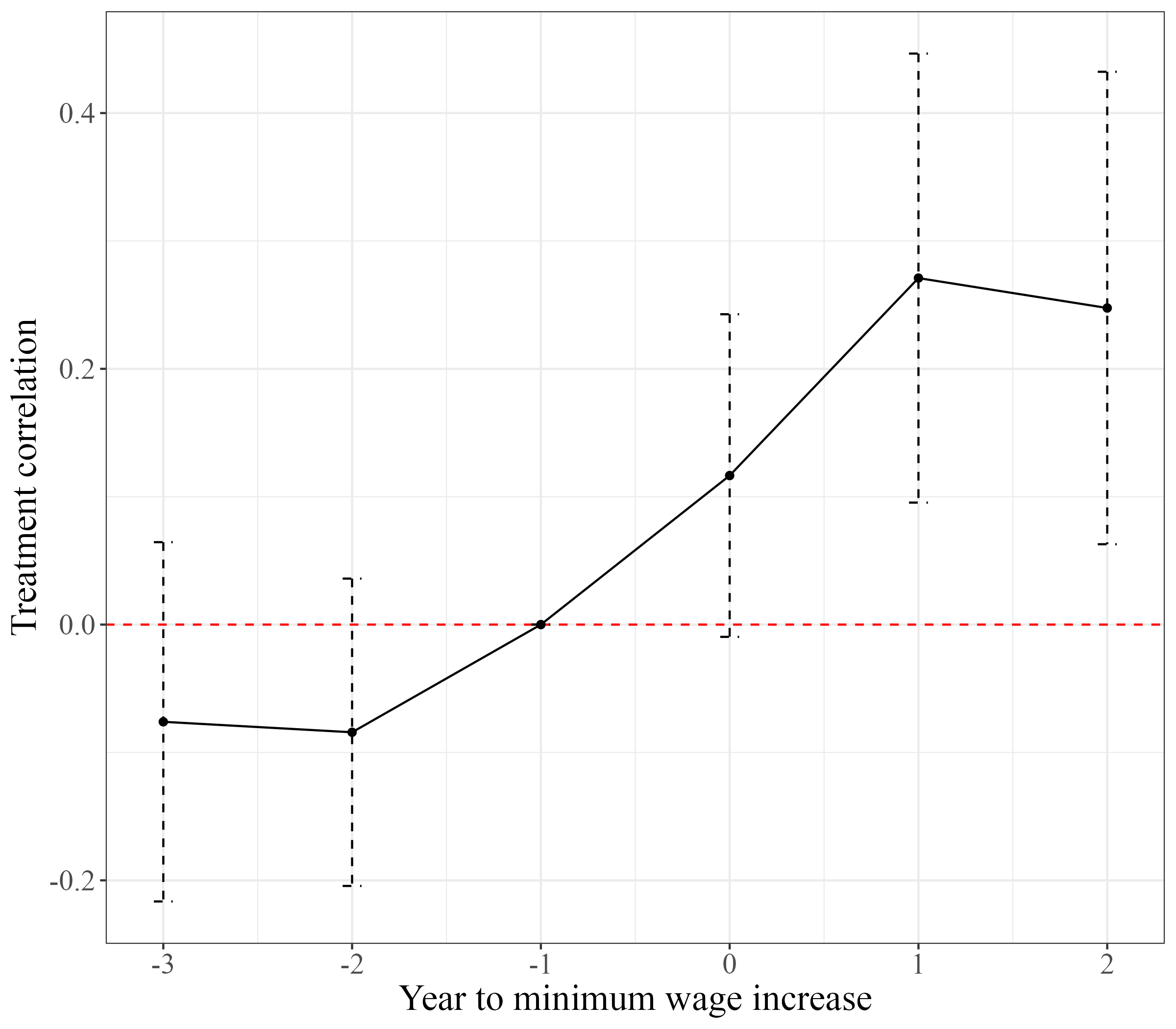} 
    \caption{Correlation Between Minimum Wage Increase and Medicaid Expansion}
    \label{fig:diag}

    \fnote{This figure reports the estimates of the confoundedness between minimum wage increase ACA Medicaid Expansion, measured by the diagnostics in equation \ref{eq:diag} using never-treated states as control. The points are estimates for the relative size of the bias in the unconditional estimator by the effect from the expansion, and the bars are 95\% confidence interval. Standard errors are clustered at the state level.}
\end{figure}

Figure \ref{fig:diag} reports the correlation between minimum wage increase and Medicaid expansion using the diagnostic in Equation \ref{eq:diag}. States that increased the minimum wage also tend to adopt the Medicaid Expansion. For example. in a staggered DiD design with unconditional estimators, the estimates for the first year after the increase have a bias with the size of \getvar{bias_short}\% of the Medicaid Expansion effect. Overall, the average bias across all post-periods is \getvar{bias_avg}\% the Medicaid Expansion effect. Alarmingly, the correlation was not significant before the minimum wage increase, suggesting that pre-trends would not have been a good diagnostic in this case. 

\bigskip

\begin{figure}
    \centering
    \includegraphics[width=0.9\textwidth]{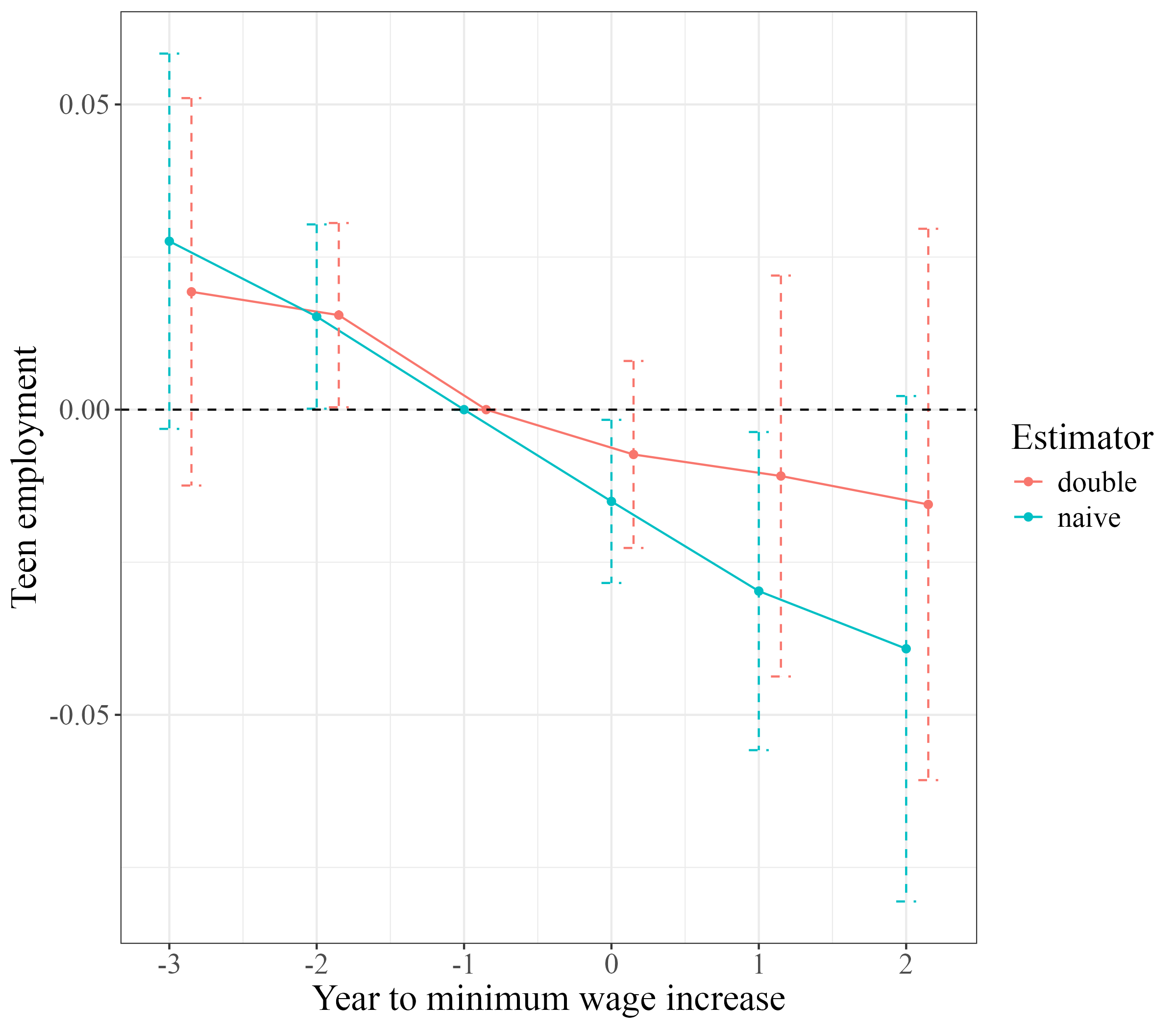} 
    \caption{Estimates of Minimum Wage Increase on Teen Employment}
    \label{fig:est}
    \fnote{This figure reports the estimates of the dynamic effect of the minimum wage increase on teen employment, using never-treated states as control. The points are estimated average treatment effect and the bars cover the 95\% confidence interval. The blue lines (``naive'') report the result with unconditional staggered differences in difference estimators. The red lines report the result with the double DiD estimators. Standard errors are clustered at the state level.}
\end{figure}

Figure \ref{fig:est} reports the short-term effects of minimum wage on teen employment, I focus on the year after the increase because treatment in the middle of the year attenuates the effect in the first year. With a staggered DiD with CS' unconditional estimator (``naive'' specification), I find negative effects on teen employment, similar to CS and \cite{Dube2016-pr}. However, when I use the Double DiD estimator to control for the expansion (``double'' specification), the negative effect in the first year after the increase weakens from \getvar{short_naive} p.p. to \getvar{short_double} p.p. and turns statistically insignificant. Summarizing all post-increase periods, the overall effect changed from \getvar{post_naive} p.p. to \getvar{post_double} p.p., but both are insignificant. This suggests that the negative effect of the ``naive'' specification is spurious due to the omitted event bias from the Medicaid Expansion. 


\bigskip

\begin{figure}
    \centering
    \includegraphics[width=0.9\textwidth]{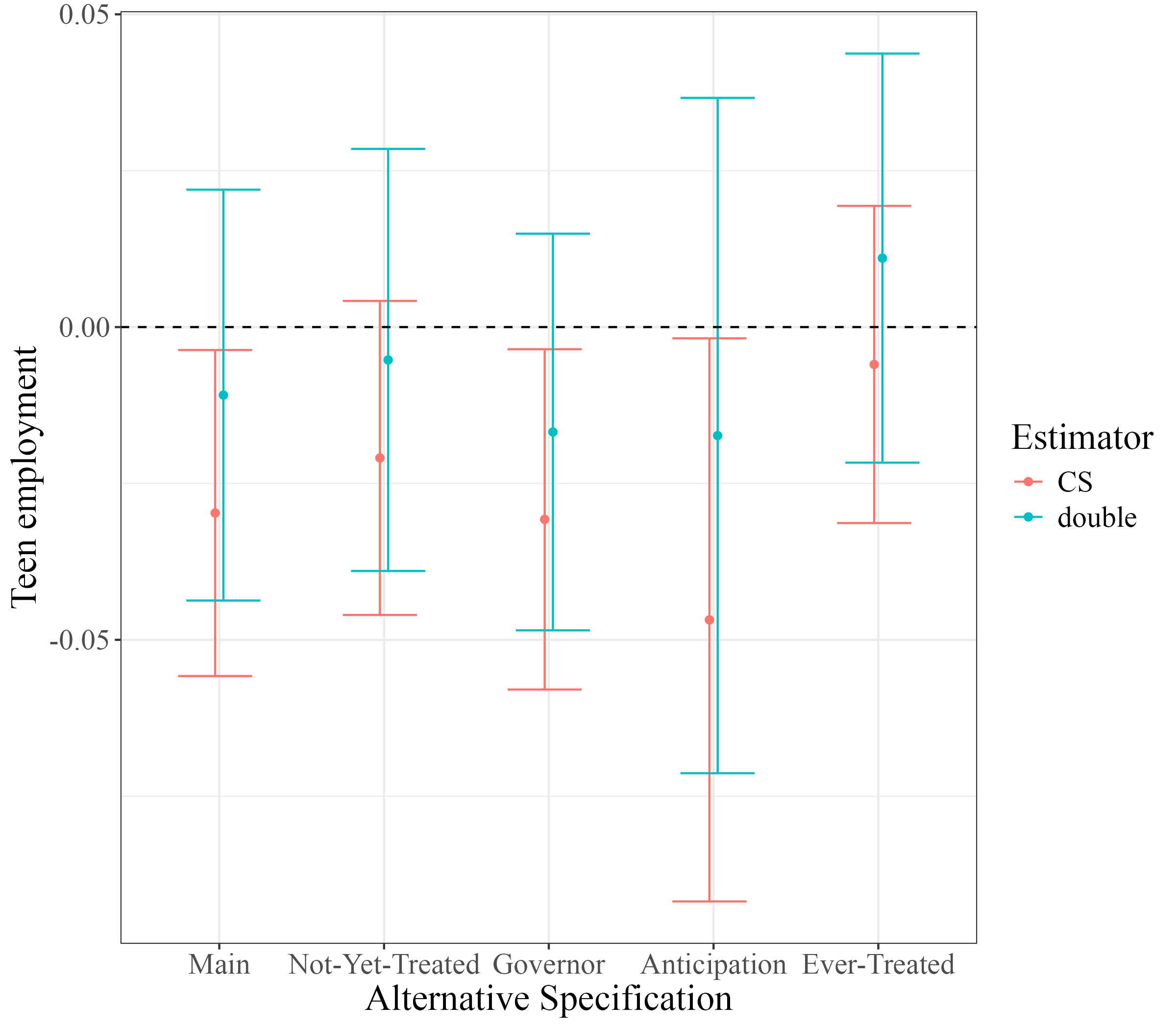} 
    \caption{Estimates of Minimum Wage Increase on Teen Employment under Alternative Specifications}
    \label{fig:robust}
    \fnote{This figure reports the estimates of the effect of the minimum wage increase on teen employment in the first year after the increase across specifications. ``Main' is the specification in Figure \ref{fig:est}. ``Not-Yet-Treated'' added not yet treated units as controls. ``Governor'' controls the party of the governor in 2010 with inverse probability weight. ``Anticipation'' allows for 1 year of treatment anticipation. ``Ever-Treated'' removes never-treated states as controls. The points are estimated average treatment effect and the bars cover the 95\% confidence interval. The blue lines (``naive'') report the result with unconditional staggered differences in difference estimators. The red lines report the result with the double DiD estimators. Standard errors are clustered at the state level.}
\end{figure}

Figure \ref{fig:robust} shows that this negative bias is robust to alternative specifications. Allowing for 1 year of anticipation, and adding not-yet-treated units as control does not change the result qualitatively, albeit the statistical significance fluctuates due to the limited sample size at the state-year level. Interestingly, controlling for whether the Governor is a Democrat in 2010 does not change the result at all, suggesting that the confoundedness goes beyond the ruling party of the state. Finally, switching to using only ever-treated units substantially weakens both estimates, suggesting that the confoundedness is most severe among states that never increased minimum wages nor expanded Medicaid. 

\bigskip

This spurious negative effect of minimum wage on teen employment does not necessarily contradict previous empirical findings. Aside from using different periods, the empirical literature is aware of the potential bias in an unconditional staggered Difference in Difference estimate \citep{Allegretto2017-am}, and often controls for many factors in their specification. This is very likely to weaken the strength of the omitted event bias. Additionally, other unaccounted state policies can confound the Medicaid Expansion as well, creating more bias. Therefore, my takeaway from this empirical exercise is that in an important application of the DiD design, the omitted event bias matters.

\section{Discussion}

In this paper, I show confounding treatments with correlated timing create bias in the standard staggered DiD estimators and provide a solution that allows for rich treatment effect heterogeneity. To eliminate this omitted event bias, I address the problem of using confounded units as controls by using only neither-yet-treated units. For the problem of a confounded treated group, I propose a second DiD design that employs a parallel treatment effect assumption. I also show that for the effect of minimum wage on teen employment, the Medicaid Expansion creates a spurious negative effect. Work for future research includes incorporating other advancements in the new DiD literature, a conditional parallel trend assumption for the treatment effect. 

\FloatBarrier


\pagebreak
\bibliographystyle{aea}
\bibliography{main}

@ARTICLE{De_Chaisemartin2020-ke,
  title     = "Two-Way Fixed Effects Estimators with Heterogeneous Treatment
               Effects",
  author    = "de Chaisemartin, Clément and D'Haultfœuille, Xavier",
  journal   = "Am. Econ. Rev.",
  publisher = "aeaweb.org",
  volume    =  110,
  number    =  9,
  pages     = "2964--2996",
  month     =  sep,
  year      =  2020
}

@ARTICLE{Fricke2017-jo,
  title     = "Identification based on difference-in-differences approaches with
               multiple treatments",
  author    = "Fricke, Hans",
  journal   = "Oxf. Bull. Econ. Stat.",
  publisher = "Wiley",
  volume    =  79,
  number    =  3,
  pages     = "426--433",
  month     =  jun,
  year      =  2017,
  language  = "en"
}

@ARTICLE{King1992-mv,
  title   = "\textit{New Jersey Sales Tax Cut 1 Cent As G.O.P. Overrides Florio
             Veto}",
  author  = "King, Wayne",
  journal = "New York Times",
  year    =  1992
}

@ARTICLE{Murphy2001-ha,
  title     = "Marginal mean models for dynamic regimes",
  author    = "Murphy, S A and van der Laan, M J and Robins, J M and {CPPRG}",
  journal   = "J. Am. Stat. Assoc.",
  publisher = "Informa UK Limited",
  volume    =  96,
  number    =  456,
  pages     = "1410--1423",
  month     =  dec,
  year      =  2001,
  language  = "en"
}

@ARTICLE{Card1993-ej,
  title     = "Minimum wages and employment: A case study of the fast food
               industry in New Jersey and Pennsylvania",
  author    = "Card, D and Krueger, A B",
  publisher = "nber.org",
  year      =  1993
}

@ARTICLE{Allegretto2017-am,
  title     = "Credible research designs for minimum wage studies: A response to
               Neumark, Salas, and Wascher",
  author    = "Allegretto, Sylvia and Dube, Arindrajit and Reich, Michael and
               Zipperer, Ben",
  journal   = "Ind. Labor Relat. Rev.",
  publisher = "SAGE Publications",
  volume    =  70,
  number    =  3,
  pages     = "559--592",
  month     =  may,
  year      =  2017,
  language  = "en"
}

@ARTICLE{Cengiz2019-vr,
  title     = "The effect of minimum wages on low-wage jobs",
  author    = "Cengiz, Doruk and Dube, Arindrajit and Lindner, Attila and
               Zipperer, Ben",
  journal   = "Q. J. Econ.",
  publisher = "Oxford University Press (OUP)",
  volume    =  134,
  number    =  3,
  pages     = "1405--1454",
  month     =  aug,
  year      =  2019,
  language  = "en"
}

@ARTICLE{de-Chaisemartin2023-jb,
  title     = "Two-way fixed effects and differences-in-differences estimators
               with several treatments",
  author    = "de Chaisemartin, Clément and D'Haultfœuille, Xavier",
  journal   = "J. Econom.",
  publisher = "Elsevier BV",
  volume    =  236,
  number    =  2,
  pages     =  105480,
  month     =  oct,
  year      =  2023,
  language  = "en"
}

@ARTICLE{Sorens2008-ek,
  title     = "{U}.{S}. state and local public policies in 2006: A new database",
  author    = "Sorens, Jason and Muedini, Fait and Ruger, William P",
  journal   = "State Politics Policy Q.",
  publisher = "Cambridge University Press (CUP)",
  volume    =  8,
  number    =  3,
  pages     = "309--326",
  year      =  2008,
  language  = "en"
}

@ARTICLE{Matthay2022-tx,
  title     = "The revolution will be hard to evaluate: How co-occurring policy
               changes affect research on the health effects of social policies",
  author    = "Matthay, Ellicott C and Hagan, Erin and Joshi, Spruha and Tan,
               May Lynn and Vlahov, David and Adler, Nancy and Glymour, M Maria",
  journal   = "Epidemiol. Rev.",
  publisher = "Oxford University Press (OUP)",
  volume    =  43,
  number    =  1,
  pages     = "19--32",
  month     =  jan,
  year      =  2022,
  language  = "en"
}

@ARTICLE{Goodman-Bacon2021-ze,
  title     = "Difference-in-differences with variation in treatment timing",
  author    = "Goodman-Bacon, Andrew",
  journal   = "J. Econom.",
  publisher = "Elsevier",
  volume    =  225,
  number    =  2,
  pages     = "254--277",
  month     =  dec,
  year      =  2021
}

@ARTICLE{Roth2022-cb,
  title         = "What's Trending in Difference-in-Differences? A Synthesis of
                   the Recent Econometrics Literature",
  author        = "Roth, Jonathan and Sant'Anna, Pedro H C and Bilinski, Alyssa
                   and Poe, John",
  journal       = "arXiv [econ.EM]",
  month         =  jan,
  year          =  2022,
  archivePrefix = "arXiv",
  primaryClass  = "econ.EM"
}

@ARTICLE{Callaway2021-vo,
  title     = "Difference-in-differences with multiple time periods",
  author    = "Callaway, B and Sant'Anna, P H C",
  journal   = "J. Econom.",
  publisher = "Elsevier",
  year      =  2021
}

@ARTICLE{Griffin2023-nu,
  title     = "Methodological considerations for estimating policy effects in
               the context of co-occurring policies",
  author    = "Griffin, Beth Ann and Schuler, Megan S and Pane, Joseph and
               Patrick, Stephen W and Smart, Rosanna and Stein, Bradley D and
               Grimm, Geoffrey and Stuart, Elizabeth A",
  journal   = "Health Serv. Outcomes Res. Methodol.",
  publisher = "Springer Science and Business Media LLC",
  volume    =  23,
  number    =  2,
  pages     = "149--165",
  year      =  2023,
  language  = "en"
}

@ARTICLE{Roller2023-zy,
  title    = "Differences-in-differences with multiple treatments under control",
  author   = "Roller, Marcus and Steinberg, D",
  year     =  2023
}

@ARTICLE{Abadie2023-yl,
  title     = "When should you adjust standard errors for clustering?",
  author    = "Abadie, Alberto and Athey, Susan and Imbens, Guido W and
               Wooldridge, Jeffrey M",
  journal   = "Q. J. Econ.",
  publisher = "Oxford University Press",
  volume    =  138,
  number    =  1,
  pages     = "1--35",
  year      =  2023
}

@ARTICLE{Hull2018-mv,
  title         = "Estimating treatment effects in mover designs",
  author        = "Hull, Peter",
  journal       = "arXiv [econ.EM]",
  month         =  apr,
  year          =  2018,
  archivePrefix = "arXiv",
  primaryClass  = "econ.EM"
}

@ARTICLE{Caetano2024-dn,
  title         = "Difference-in-differences when parallel trends holds
                   conditional on covariates",
  author        = "Caetano, Carolina and Callaway, Brantly",
  journal       = "arXiv [econ.EM]",
  month         =  jun,
  year          =  2024,
  archivePrefix = "arXiv",
  primaryClass  = "econ.EM"
}

@ARTICLE{Neumark2006-re,
  title     = "Minimum wages and employment",
  author    = "Neumark, David and Wascher, William L",
  journal   = "Found. Trends® Microecon.",
  publisher = "Now Publishers",
  volume    =  3,
  number    = "1-2",
  pages     = "1--182",
  year      =  2006,
  language  = "en"
}

@ARTICLE{Dube2016-pr,
  title     = "Minimum wage shocks, employment flows, and labor market frictions",
  author    = "Dube, Arindrajit and Lester, T William and Reich, Michael",
  journal   = "J. Labor Econ.",
  publisher = "University of Chicago Press",
  volume    =  34,
  number    =  3,
  pages     = "663--704",
  month     =  jul,
  year      =  2016,
  language  = "en"
}

@MISC{KFF2024-cb,
  title        = "Status of State Medicaid Expansion Decisions",
  author       = "{KFF}",
  booktitle    = "Kaiser Family Foundation",
  year         =  2024,
  howpublished = "\url{https://www.kff.org/affordable-care-act/issue-brief/status-of-state-medicaid-expansion-decisions-interactive-map/}",
  note         = "Accessed: 2024--"
}

@MISC{University-of-Kentucky-Center-for-Poverty-Research2024-su,
  title  = "{UKCPR} National Welfare Data, 1980-2022",
  author = "{University of Kentucky Center for Poverty Research}",
  year   =  2024,
  note   = "Title of the publication associated with this dataset: UKCPR
            National Welfare Data, 1980-2022"
}

@INCOLLECTION{Abowd2009-ul,
  title     = "5. The {LEHD} Infrastructure Files and the Creation of the
               Quarterly Workforce Indicators",
  author    = "Abowd, John M and Stephens, Bryce E and Vilhuber, Lars and
               Andersson, Fredrik and McKinney, Kevin L and Roemer, Marc and
               Woodcock, Simon",
  booktitle = "Producer Dynamics",
  publisher = "University of Chicago Press",
  pages     = "149--234",
  month     =  may,
  year      =  2009,
  language  = "en"
}

@ARTICLE{Khan2010-pu,
  title     = "Irregular identification, support conditions, and inverse weight
               estimation",
  author    = "Khan, S and Tamer, E",
  journal   = "Econometrica",
  publisher = "Wiley Online Library",
  year      =  2010
}
\pagebreak
\begin{appendices}
\section{Proofs}\label{app:proof}

\subsection{Proof of Theorem \ref{thm:gtbias}}

Within this proof, I denote the base period by $b \equiv g^1-1$. Let's first decompose the moment into four parts

\begin{align*}
 ATT^{CS}(g^1, t) \equiv &  \mathbb{E}[Y_t - Y_b|G^1_{g^1} = 1] - \mathbb{E}[Y_t - Y_b|D^1_t = 0] \\
= & \mathbb{E}[Y_t|G^1_{g^1} = 1] - \mathbb{E}[Y_b|G^1_{g^1} = 1]  - \mathbb{E}[Y_t|D^1_t = 0] + \mathbb{E}[Y_b|D^1_t = 0].
\end{align*}

For the treated group in $t$, 

\begin{align*}
\mathbb{E}[Y_t|G^1_{g^1} = 1]  = & \mathbb{E}[Y_t|G^1_{g^1} = 1, D^2_t = 1]P(D^2_t = 1 | G^1_{g^1} = 1) + \mathbb{E}[Y_t|G^1_{g^1} = 1, D^2_t = 0]P(D^2_t = 0| G^1_{g^1} = 1) \\
= & \mathbb{E}[Y_t(g^1, G^2)|G^1_{g^1} = 1, D^2_t = 1]P(D^2_t = 1 | G^1_{g^1} = 1) \\
& + \mathbb{E}[Y_t(g^1, 0)|G^1_{g^1} = 1, D^2_t = 0]P(D^2_t = 0| G^1_{g^1} = 1) \\
= & \mathbb{E}[Y_t(g^1, 0)|G^1_{g^1} = 1] + \mathbb{E}[Y_t(g^1, G^2)-Y_t(g^1, 0)|G^1_{g^1} = 1, D^2_t = 1]P(D^2_t = 1 | G^1_{g^1} = 1) \\
= & \mathbb{E}[Y_t(g^1, 0)|G^1_{g^1} = 1] + (ATT - ATT^1)P(D^2_t = 1 | G^1_{g^1} = 1).
\end{align*}

For the treated group in $b$, 

\begin{align*}
 \mathbb{E}[Y_b|G^1_{g^1} = 1] = &  \mathbb{E}[Y_b|G^1_{g^1} = 1, D^2_b = 1]P(D^2_b = 1 | G^1_{g^1} = 1) + \mathbb{E}[Y_b|G^1_{g^1} = 1, D^2_b = 0]P(D^2_b = 0| G^1_{g^1} = 1) \\
= &  \mathbb{E}[Y_b(0, G^2)|G^1_{g^1} = 1, D^2_b = 1]P(D^2_b = 1 | G^1_b = 1) \\
& + \mathbb{E}[Y_b(0, 0)|G^1_{g^1}= 1, D^2_b = 0]P(D^2_b = 0| G^1_{g^1} = 1) \\
= & \mathbb{E}[Y_b(0, 0)|G^1_{g^1} = 1] + \mathbb{E}[Y_b(0, G^2)-Y_b(0, 0)|G^1_{g^1} = 1, D^2_b = 1]P(D^2_b = 1| G^1_{g^1} = 1) \\
= & \mathbb{E}[Y_b(0, 0)|G^1_{g^1} = 1] + ATT^2P(D^2_b = 1| G^1_{g^1} = 1).
\end{align*}

For the control group in $t$

\begin{align*}
\mathbb{E}[Y_t|D^1_t = 0] = & \mathbb{E}[Y_t|D^1_t = 0, D^2_t = 1]P(D^2_t = 1 | D^1_t = 0 ) + \mathbb{E}[Y_t|D^1_t = 0, D^2_t = 0]P(D^2_t = 0|  D^1_t = 0) \\
= & \mathbb{E}[Y_t(0, G^2)|D^1_t = 0, D^2_t = 1]P(D^2_t = 1 | D^1_t = 0 ) \\
& +  \mathbb{E}[Y_t(0,0)|D^1_t = 0, D^2_t = 0]P(D^2_t = 0|  D^1_t = 0) \\ 
= & \mathbb{E}[Y_t(0, 0)|D^1_t = 0] +\mathbb{E}[Y_t(0, G^2)-Y_t(0, 0)|D^1_t = 0, D^2_t = 1]P(D^2_t = 1 | D^1_t = 0 ) \\
= & \mathbb{E}[Y_t(0, 0)|D^1_t = 0] + ATT^2P(D^2_t = 1 | D^1_t = 0 ).
\end{align*}

For the control group in $b$

\begin{align*}
\mathbb{E}[Y_b|D^1_t = 0] = & \mathbb{E}[Y_b|D^1_t = 0, D^2_b = 1]P(D^2_b = 1 | D^1_t = 0) + \mathbb{E}[Y_b|D^1_t = 0, D^2_b = 0]P(D^2_b = 0| D^1_t = 0) \\
= &  \mathbb{E}[Y_b(0, g^2)|D^1_t = 0, D^2_b = 1]P(D^2_b = 1 | D^1_t = 0)\\
& + \mathbb{E}[Y_b(0,0)|D^1_t = 0, D^2_b = 0]P(D^2_b = 0| D^1_t = 0) \\
= & \mathbb{E}[Y_b(0, 0)|D^1_t = 0] + \mathbb{E}[Y_b(0, g^2)-Y_b(0, 0))|D^1_t = 0, D^2_b = 1]P(D^2_b = 1| D^1_t = 0)   \\
= & \mathbb{E}[Y_b(0, 0)|D^1_t = 0] +ATT^2P(D^2_b = 1| D^1_t = 0).
\end{align*}

Combine the four parts, 

\begin{align*}
 ATT^{CS}(g^1, t) = & \mathbb{E}[Y_t(g^1, 0)|G^1_{g^1} = 1]  -  \mathbb{E}[Y_b(0, 0)|G^1_{g^1} = 1] - \mathbb{E}[Y_t(0, 0)|D^1_t = 0]  + \mathbb{E}[Y_b(0, 0)|D^1_t = 0] \\
 & (ATT - ATT^1)P(D^2_t = 1 | G^1_{g^1} = 1) \\
 &-
ATT^2 P(D^2_b = 1| G^1_{g^1} = 1) \\ 
& -
ATT^2P(D^2_t = 1 | D^1_t = 0 ) + 
ATT^2P(D^2_b = 1| D^1_t = 0)  \big\} \\
= & ATT_{10}(g^1, t) \\
    & +ATT^2\left(P(D^2_t = 1, D^2_b = 0 | G^1_{g^1 }= 1) -P(D^2_t = 1, D^2_b = 0 | D^1_t = 0 )\right) \\
    & + (ATT-ATT^1-ATT^2) P(D^2_t = 1 | G^1_{g^1} = 1).
\end{align*}

\subsection{Proof of Lemma \ref{lem:assequiv}}

In the following parts, I abbreviate CS's assumption X as CSX and my assumption X as AX. First, see that $Y_t(g,g) = Y_t(g)$ since if $G^1_i = G^2_i = g$, $Y_t(g) = Y_{i}(G^1, G^2) = Y_{t}(g,g)$, else $Y_t(g,g) = Y_t(g)$. Also $D_{t} = 1-(1-D^1_t)(1-D^2_t)$ since $D_t \equiv \mathbbm{1}\{t \geq G_i\} = \mathbbm{1}\{t \geq \min(G^1_i, G^2_i)\} = 1-(1-\mathbbm{1}\{t \geq G^1_i\})(1-\mathbbm{1}\{t \geq G^2_i\}) = 1-(1-D^1_t)(1-D^2_t)$. 

\bigskip

For CS1 (Irreversibility of Treatment)),  $D_{t-1}=1$ implies either $D^1_{t-1} = 1$ or $D^2_{t-1} = 1$. Hence either $D^1_{t} = 1$ or $D^2_{1} = 1$ (A\ref{ass:irrtreat}), and $D_t = 1-(1-D^1_t)(1-D^2_t) = 1-0 = 1.$

\bigskip

For CS2 (Random Sampling), $Y, X$ are the same since we only subsample by units. $D^1_t, D^2_t$ are i.i.d (A\ref{ass:sample}) implies that $1-(1-D^1_t)(1-D^2_t) = D_t$ is i.i.d.

\bigskip

For CS3 (Limited Treatment Anticipation) with $\delta = 0$, take $t,g$ such that $t<g$

\begin{align*}
    E[Y_t(g)|X, G_g = 1] = & E[Y_t(g,g)|X, G_g = 1, \min(G^1, G^2) \neq g](1-P( \min(G^1, G^2) = g|X, G_g = 1)) \\ 
    & + E[Y_t(G^1,G^2)|X, G_g = 1, g = \min(G^1, G^2) ]P( \min(G^1, G^2) = g|X, G_g = 1) \\
    = & E[Y_t(0,0)|X, G_g = 1, \min(G^1, G^2) \neq g](1-P( \min(G^1, G^2) = g|X, G_g = 1)) \\ 
    & + E[Y_t(0,0)|X, G_g = 1, g = \min(G^1, G^2) ]P( \min(G^1, G^2) = g|X, G_g = 1) && \text{A\ref{ass:anticipation}}\\
    = & E[Y_t(0,0)|X, G_g = 1] \\
    = & E[Y_t(0)|X, G_g = 1].
\end{align*}

For CS4 (Conditional Parallel Trends Based on a ``Never-Treated'' Group), take $t,g$ such that $t\geq g$

\begin{align*}
    E[Y_t(0)- Y_{t-1}(0)|X, G_g = 1] = & \mathbb{E}[Y_t(0,0)- Y_{t-1}(0,0)|X, \min(G^1, G^2) = g] \\
    = & \sum^T_{g^1 = 2}\sum^T_{g^2 = 2}\mathbbm{1}\{\min(G^1, G^2) = g\}P(G^1_{g^1} = 1, G^2_{g^2} = 1|X, \min(G^1, G^2) = g)\\
    & \mathbb{E}[Y_t(0,0)- Y_{t-1}(0,0)|X, G^1_{g^1} = 1, G^2_{g^2} = 1]\\
    = & \sum^T_{g^1 = 2}\sum^T_{g^2 = 2}\mathbbm{1}\{\min(G^1, G^2) = g\}P(G^1_{g^1} = 1, G^2_{g^2} = 1|X, \min(G^1, G^2) = g)\\
    & \mathbb{E}[Y_t(0, 0) - Y_{t-1}(0, 0)|X, D^1_s = 0, D^2_s = 0, G^1_{\infty} = 1, G^2_{\infty} = 1] && \text{A\ref{ass:pt-never}} \\
     & \because G^1_{\infty} = 1, G^2_{\infty} = 1 \implies G = \min(G^1, G^2) = \infty, \\
     & G = \infty \implies D^1_s = 0, D^2_s = 0 \\
    = &  \sum^T_{g^1 = 2}\sum^T_{g^2 = 2}\mathbbm{1}\{\min(G^1, G^2) = g\}P(G^1_{g^1} = 1, G^2_{g^2} = 1|X, \min(G^1, G^2) = g) \\
    & \mathbb{E}[Y_t(0, 0) - Y_{t-1}(0, 0)|X, G_{\infty} = 1] \\
    = &  \mathbb{E}[Y_t(0) - Y_{t-1}(0)|X, G_{\infty} = 1].
\end{align*}

For CS5 (Conditional Parallel Trends Based on a ``Not-Yet-Treated'' Group), take $t,g$ such that $t\geq g$ and $t \leq s < \hat{g}$

\begin{align*}
    E[Y_t(0)- Y_{t-1}(0)|X, G_g = 1] = & \mathbb{E}[Y_t(0,0)- Y_{t-1}(0,0)|X, \min(G^1, G^2) = g] \\
    = & \sum^T_{g^1 = 2}\sum^T_{g^2 = 2}\mathbbm{1}\{\min(G^1, G^2) = g\}P(G^1_{g^1} = 1, G^2_{g^2} = 1|X, \min(G^1, G^2) = g)\\
    & \mathbb{E}[Y_t(0,0)- Y_{t-1}(0,0)|X, G^1_{g^1} = 1, G^2_{g^2} = 1]\\
    = & \sum^T_{g^1 = 2}\sum^T_{g^2 = 2}\mathbbm{1}\{\min(G^1, G^2) = g\}P(G^1_{g^1} = 1, G^2_{g^2} = 1|X, \min(G^1, G^2) = g)\\
    & \mathbb{E}[Y_t(0, 0) - Y_{t-1}(0, 0)|X, D^1_s = 0, D^2_s = 0, G^1_{g^1} = 0, G^2_{g^2} = 0] && \text{A\ref{ass:pt}} \\
     & \because D^1_s = 0, D^2_s = 0 \implies D_s = 0, G^1_{g^1} = 0, G^2_{g^2} = 0, \\
     & \forall i \: s.t. \min(G^1_i, G^2_i) = g, G^1_i \neq, g^1, G^2_i \neq, g^2 \implies \forall i, G_i \neq g.\\
    = &  \sum^T_{g^1 = 2}\sum^T_{g^2 = 2}\mathbbm{1}\{\min(G^1, G^2) = g\}P(G^1_{g^1} = 1, G^2_{g^2} = 1|X, \min(G^1, G^2) = g) \\
    & \mathbb{E}[Y_t(0, 0) - Y_{t-1}(0, 0)|X, D_s = 0, G_{g} = 0] \\
    = &  \mathbb{E}[Y_t(0) - Y_{t-1}(0)|X, D_s = 0, G_{g} = 0].
\end{align*}

Finally, for CS6, take $g$ such that there exists $g^1, g^2 \in G$ such that $g = \min(g^1, g^2)$

\begin{align*}
    P(G_g = 1) =  & \sum^T_{g^1 = 2}\sum^T_{g^2 = 2}\mathbbm{1}\{\min(G^1, G^2) = g\}P(G^1_{g^1} = 1, G^2_{g^2} = 1) \\
    >  &  \sum^T_{g^1 = 2}\sum^T_{g^2 = 2}\mathbbm{1}\{\min(G^1, G^2) = g\}\epsilon_{g^1g^2t} > 0 && \text{A\ref{ass:overlap}}\\
\end{align*}

and 

\begin{align*}
    p_{g,t}(X) =  & P(G_g = 1|X, G_g + (1-D_t)(1-G_g)= 1) \\
    = & P(G_g = 1|X, G \geq g) \\
    = & \sum^T_{g^1 = 2}\sum^T_{g^2 = 2}\mathbbm{1}\{\min(G^1, G^2) \geq g\} \\
   & P(G^1_{g^1} = 1, G^2_{g^2} = 1|X, G^1 \geq g, G^2 \geq g) \\
    < & \sum^T_{g^1 = 2}\sum^T_{g^2 = 2}\mathbbm{1}\{\min(G^1, G^2) \geq g\} 1-\epsilon_{g^1g^2t}  < 1. && \text{A\ref{ass:overlap}}\\
\end{align*}

\subsection{Proof of Theorem \ref{thm:ggtident}}

Note that  $G_{\min(g^1, g^2)} = 1, W(g^1, g^2) = 1 \iff G^1 = 1, G^2 = 1$. Also, $G_{\min(g^1, g^2)}+(1-D_t)(1- G_{\min(g^1, g^2)}) = 1, W(g^1, g^2) = 1 \iff (1-D_t)(1-G^1_{g^1}G^2_{g^2}) = 1$. And $G_{\min(g^1, g^2)} = 0, W(g^1, g^2) = 1 \iff G^1 = 0, G^2 = 0$.

\bigskip

We want to show that in the subsample $\mathbb{I}(g^1, g^2)$, my identifying moments are equivalent to CS's moments. Let's start with the unconditional version. 

\begin{claim}

In the subsample $\mathbb{I}(g^1, g^2)$,

$$ATT_{unc}(g^1, g^2, t) =ATT^{ny,CS}_{unc}(\min(g^1, g^2), t ;0),$$

where 

$$ATT^{ny,CS}_{unc}(g,t;\delta) = \mathbb{E}[Y_t-Y_{g-\delta-1}|G_g = 1] - \mathbb{E}[Y_t-Y_{g-\delta-1}|D_{d+\delta} = 0]$$

as in CS equation 2.9.

\bigskip

Proof: 

\begin{align*}
    & ATT^{ny, CS}_{unc}(\min(g^1, g^2), t;0) \\
    = & \mathbb{E}[Y_t - Y_{\min(g^1, g^2)-1} | G_{\min(g^1, g^2)} = 1] \\
    & - \mathbb{E}[Y_t - Y_{\min(g^1, g^2)-1} | D_{t} = 0]
 && \text{CS Eq 2.9} \\
 = & \mathbb{E}[Y_t - Y_{\min(g^1, g^2)-1} | G_{\min(g^1, g^2)} = 1, W(g^1, g^2) = 1] \\
    & - \mathbb{E}[Y_t - Y_{\min(g^1, g^2)-1} | D_{t} = 0, W(g^1, g^2) = 1]
 && i\in \mathbb{I}(g^1, g^2) \\
= & \mathbb{E}[Y_t - Y_{\min(g^1, g^2)-1}|G^1_{g^1} = 1, G^2_{g^2} = 1] - \mathbb{E}[Y_t - Y_{\min(g^1, g^2)-1}|D_t = 0] \\
    = & \mathbb{E}[Y_t - Y_{\min(g^1, g^2)-1}|G^1_{g^1} = 1, G^2_{g^2} = 1]- \mathbb{E}[Y_t - Y_{\min(g^1, g^2)-1}|D^1_t = 0, D^2_t = 0] \\
    = & ATT_{unc}(g^1, g^2, t). 
\end{align*}

\end{claim}

\bigskip

Then, the inverse probability weighting version. 

\begin{claim}

Suppose that $i \in \mathbb{I}(g^1, g^2)$,

$$ATT_{ipw}(g^1, g^2, t) = ATT^{ny,CS}_{ipw}(\min(g^1, g^2), t ;0),$$

where $ATT^{ny,CS}_{ipw}(g,t;\delta)$ is defined as in CS equation 2.5.

\bigskip

Proof: For the propensity score  $p^{CS}_{g, t}(X) \equiv P(G_{g} = 1|X, G_{g}+(1-D_t)(1- G_{g}) = 1) $, we have 

\begin{align*}
    &p^{CS}_{\min(g^1, g^2), t}(X) \\
    = & P(G_{\min(g^1, g^2)} = 1|X, G_{\min(g^1, g^2)}+(1-D_t)(1- G_{\min(g^1, g^2)}) = 1, W(g^1, g^2) = 1) && i\in \mathbb{I}(g^1, g^2)\\
    = & P(G^1 = 1, G^2 = 1 |X, (1-D_t)(1-G^1_{g^1}G^2_{g^2}) = 1) \\
    = & p_{g^1, g^2,t}(x).
\end{align*}

For the inverse propensity weight $w^{comp,CS}(g,t) \equiv \frac{p^{CS}_{g,t}(1-D_t)(1-G_g)}{1-p^{CS}_{g,t}(X)}$: 

\begin{align*}
     & w^{comp,CS}(g,t) \\
     = & \frac{p^{CS}_{\min(g^1, g^2),t}(1-D_t)(1-G_{\min(g^1, g^2)})}{1-p^{CS}_{\min(g^1, g^2),t}(X)}\\
     = & \frac{p^{CS}_{\min(g^1, g^2),t}(1-D_t)(1-G^1_{g^1}G^2_{g^2})}{1-p^{CS}_{\min(g^1, g^2),t}(X)} && i \in \mathbb{I}(g^1, g^2)\\
     = & \frac{p_{g^1, g^2,t}(X)(1-D_t)(1-G^1_{g^1}G^2_{g^2})}{1-p_{g^1, g^2,t}(X)} && p^{CS}(\cdot) = p(\cdot) \\
     = &  w^{comp}(g^1, g^2, t). 
\end{align*}

Now, for inverse probability weighting moment:

\begin{align*}
& ATT^{ny,CS}_{ipw}(\min(g^1, g^2), t; 0) \\
 = &\mathbb{E} \left[ \left( \frac{G_g}{\mathbb{E}[G_g]} -  \frac{\frac{p_{g,t}(X)(1-D_{t})(1-G_g)}{1-p_{g,t}(X)} }{\mathbb{E} \left[ \frac{p_{g,t}(X)(1-D_{t})(1-G_g)}{1-p_{g,t}(X)} \right]}\right) \left( Y_t - Y_{g-1} \right) | g = \min(g^1, g^2)\right] && \text{CS Eq 2.5} \\
 = &\mathbb{E} \left[ \left( \frac{G_g}{\mathbb{E}[G_g]} -  \frac{w^{comp,CS}(g,t)}{\mathbb{E}[w^{comp,CS}(g,t)]}\right) \left( Y_t - Y_{g-1} \right) | g = \min(g^1, g^2)\right]  \\
= & \mathbb{E} \left[\left( \frac{G^1_{g^1}G^2_{g^2}}{\mathbb{E}[G^1_{g^1}G^2_{g^2}]} -  \frac{ w^{comp}(g^1, g^2, t)}{\mathbb{E} [w^{comp}(g^1, g^2, t)]} \right)\left( Y_t - Y_{\min(g^1, g^2)-1} \right)\right] && i \in \mathbb{I}(g^1, g^2)\\
= & ATT_{ipw}(g^1, g^2, t). 
\end{align*}

\end{claim}

Then, the outcome regression version. 

\begin{claim}

In the subsample $\mathbb{I}(g^1, g^2)$,

$$ATT_{or}(g^1, g^2, t) = ATT^{ny,CS}_{or}(\min(g^1, g^2), t ;0),$$

where $ATT^{ny,CS}_{or}$ is defined as in CS equation 2.6.

\bigskip

Proof: first, for the control function, $m^{CS}_{g,t}(X) \equiv \mathbb{E}[Y_t - Y_{g-1}|X, D_t = 0, G_{g} = 0]$, we have

\begin{align*}
    &m^{CS}_{\min(g^1, g^2),t}(X) \\
    = & \mathbb{E}[Y_t - Y_{\min(g^1, g^2)-1}|X, D_t = 0, G_{\min(g^1, g^2)} = 0\\
     = &  \mathbb{E}[Y_t - Y_{\min(g^1, g^2)-1}|X, D_t = 0, G^1_{g^1} = 0, G^2_{g^2} = 0] && i\in \mathbb{I}(g^1, g^2)\\
     = &  m(g^1, g^2, t),
\end{align*}

then for the identifying moment:

\begin{align*}
& ATT^{ny,CS}_{dr}(\min(g^1, g^2), t; 0) \\
& = \mathbb{E} \left[ \left( \frac{G_g}{\mathbb{E}[G_g]}\right) \left( Y_t - Y_{g-1} - m^{ny}_{g,t}(X) \right) | g = \min(g^1, g^2) \right] && \text{CS Eq 2.6} \\
= & \mathbb{E} \left[ \left( \frac{G^1_{g^1}G^2_{g^2}}{\mathbb{E}[G^1_{g^1}G^2_{g^2}]}  \right)\left( Y_t - Y_{\min(g^1, g^2)-1} - m^{ny}_{g^1, g^2,t}(X) \right) \right] && i\in \mathbb{I}(g^1, g^2) \\
= & ATT_{or}(g^1, g^2, t).
\end{align*}

\end{claim}

\bigskip

Finally, for the doubly-robust: 

\begin{claim}

In the subsample $\mathbb{I}(g^1, g^2)$,

$$ATT_{dr}(g^1, g^2, t) = ATT^{ny,CS}_{dr}(\min(g^1, g^2), t ;0),$$

where $ATT^{ny,CS}_{dr}(\min(g^1, g^2), t ;0)$ is as defined in CS equation 2.7. 

\bigskip

Proof: 

\begin{align*}
& ATT^{ny,CS}_{dr}(\min(g^1, g^2), t; 0)\\
& = \mathbb{E} \left[ \left( \frac{G_g}{\mathbb{E}[G_g]} -  \frac{\frac{p_{g,t}(X)(1-D_{t})(1-G_g)}{1-p_{g,t}(X)} }{\mathbb{E} \left[ \frac{p_{g,t}(X)(1-D_{t})(1-G_g)}{1-p_{g,t}(X)} \right]}\right) \left( Y_t - Y_{g-1} - m^{ny}_{g,t}(X) \right) | g = \min(g^1, g^2)\right] && \text{CS Eq 2.7} \\
= & \mathbb{E} \left[ \left( \frac{G^1_{g^1}G^2_{g^2}}{\mathbb{E}[G^1_{g^1}G^2_{g^2}]} -  \frac{ w^{comp}(g^1, g^2, t)}{\mathbb{E} [w^{comp}(g^1, g^2, t)]} \right)\left( Y_t - Y_{\min(g^1, g^2)-1} - m^{ny}_{g^1, g^2,t}(X) \right) \right] 
 && i \in \mathbb{I}(g^1, g^2)\\
= & ATT_{dr}(g^1, g^2, t).
\end{align*}
\end{claim}

\bigskip

With the equivalence of moments established, Theorem 1 in CS provides identification.  For $ATT_x \in \{ATT_{unc}, ATT_{ipw}, ATT_{or}, ATT_{dr}\}$:

\begin{align*}
    ATT_x(g^1, g^2, t) = & ATT_{x}^{ny, CS}(\min(g^1, g^2), t;0) && \text{Claim 1,2,3,4}\\
    = & ATT(\min(g^1, g^2),t) && \text{CS's Thm 1} \\
    = & \mathbb{E}[Y_t(\min(g^1, g^2))-Y_t(0)|G_g = 1] \\
    = & \mathbb{E}[Y_t(\min(g^1, g^2))-Y_t(0)|G_g = 1,W(g^1, g^2)=1] && i\in \mathbb{I}(g^1, g^2)\\
    = & \mathbb{E}[Y_t(\min(g^1, g^2))-Y_t(0)|G^1_{g^1} = 1, G^2_{g^2} = 1] \\
    = & \mathbb{E}[Y_t(g^1, g^2)- Y_t(0,0)|G^1_{g^1} = 1, G^2_{g^2} = 1] \\
    = & ATT(g^1, g^2, t).
\end{align*}

In the full sample $\mathbb{I}$, note that $ATT(g^1, g^2, t)|_{i \in \mathbb{I}(g^1, g^2)} = ATT(g^1, g^2, t)$ since $G^1_{g^1} = 1, G^2_{g^2} = 1 \implies W(g^1, g^2) = 1$. 

\subsection{Proof of Theorem \ref{thm:doubledid}}

For the case when $t < g^2$, we have 

\begin{align*}
    ATT^{dd}_{id}(g^1, g^2, t) = & ATT_{id}(g^1, g^2, t) && \text{Eq. \ref{eq:ddid}, } t<g^2 \\
    = & ATT(g^1, g^2, t) && \text{Theorem \ref{thm:ggtident}} \\
    = & \mathbb{E}[Y_t(g^1, g^2)-Y_t(0, 0)|G^1_{g^1} = 1, G^2_{g^2} = 1] \\
    = & \mathbb{E}[Y_t(g^1, 0)-Y_t(0, 0)|G^1_{g^1} = 1, G^2_{g^2} = 1] && \text{Assumption \ref{ass:anticipation} }, t<g^2 \\
    = & ATT^{1}(g^1, g^2, t).
\end{align*}

For the case when $t \geq g^2$, $g^1<g^2$, we can rewrite $ATT^{imp}_{id}$ as  

\begin{align}
\begin{split}
    ATT^{imp}_{id}(g^1, g^2, t) = & ATT_{id}(g^1, g^2, g^2-1) \\
    & +  \sum_{s^2 > t}\left(ATT_{id}(g^1, s^2, t) - ATT_{id}(g^1, s^2, g^2-1)\right)P(G^2_{s^2} = 1 | G^1_{g^1} = 1, D^2_t = 0).
\end{split}
\end{align}

Then for each $s^2 > t$,

\begin{align*}
    & ATT_{id}(g^1, g^2, g^2-1) + ATT_{id}(g^1, s^2, t) - ATT_{id}(g^1, s^2, g^2-1) \\
    = & ATT(g^1, g^2, g^2-1) + ATT(g^1, s^2, t) - ATT(g^1, s^2, g^2-1) && \text{Theorem \ref{thm:ggtident}} \\
    = & ATT^1(g^1, g^2, g^2-1) + ATT^1(g^1, s^2, t) - ATT^1(g^1, s^2, g^2-1) && \text{Ass. \ref{ass:anticipation},} \: s^2 > t \\
    = & ATT^1(g^1, g^2, t) - (ATT^1(g^1, g^2, t)- ATT^1(g^1, s^2,  g^2-1)) \\
    & + (ATT^1(g^1, s^2, t) - ATT^1(g^1, s^2, g^2-1)) \\
    = &  ATT^1(g^1, g^2, t) && \text{Ass. \ref{ass:ptet}}
\end{align*}

Since there exists at least one $s^2 > t$ such that $P(G^2_{s^2} = 1 | G^1_{g^1} = 1, D^2_t = 0)>0$ as $t < \bar t^2$, we have

\begin{align*}
    & ATT^{dd}_{id}(g^1, g^2, t) \\
     = & ATT^{imp}_{id}(g^1, g^2, t) && \text{Eq. \ref{eq:ddid}, } g^1<g^2 \\
    = & ATT_{id}(g^1, g^2, g^2-1) \\
    & +  \sum_{s^2 > t}\left(ATT_{id}(g^1, s^2, t) - ATT_{id}(g^1, s^2, g^2-1)\right)P(G^2_{s^2} = 1 | G^1_{g^1} = 1, D^2_t = 0) \\
    = &  \sum_{s^2 > t}P(G^2_{s^2} = 1 | G^1_{g^1} = 1, D^2_t = 0)  \\
    & \left(ATT_{id}(g^1, g^2, g^2-1) + ATT_{id}(g^1, s^2, t) - ATT_{id}(g^1, s^2, g^2-1)\right) \\
    = &  ATT^1(g^1, g^2, t).
\end{align*}

Finally, for the case when $t \geq g^2$, $g^1>g^2$, we can rewrite Equation \ref{eq:did} as

\begin{align}
\begin{split}
    ATT^{did}(g^1, g^2, t) \equiv & ATT_{id}(g^1, g^2, t) - ATT_{id}(g^1, g^2, g^1-1) \\ 
    & - \sum_{s^1 > t}\left(ATT_{id}(s^1, g^2, t) - ATT_{id}(s^1, g^2, g^1-1)\right)P(G^1_{s^1} = 1 | G^2_{g^2} = 1, D^1_t = 0).
\end{split}
\end{align}

Then for each $s^1 > t$ ,

\begin{align*}
    & ATT_{id}(g^1, g^2, t) - ATT_{id}(g^1, g^2, g^1-1) - ATT_{id}(s^1, g^2, t) + ATT_{id}(s^1, g^2, g^1-1) \\
    = & ATT(g^1, g^2, t) - ATT(g^1, g^2, g^1-1) - ATT(s^1, g^2, t) + ATT(s^1, g^2, g^1-1) && \text{Theorem \ref{thm:ggtident}} \\
    = & ATT(g^1, g^2, t) - ATT^2(g^1, g^2, g^1-1) - ATT^2(s^1, g^2, t) + ATT^2(s^1, g^2, g^1-1)  && \text{Ass. \ref{ass:anticipation}}, s^1 > t \\
    = & ATT^1(g^1, g^2, t) + ATT^2(g^1, g^2, t) - ATT^2(g^1, g^2, g^1-1) \\
    & - ATT^2(s^1, g^2, t) + ATT^2(s^1, g^2, g^1-1) && \text{Ass. \ref{ass:addeff}} \\
    = & ATT^1(g^1, g^2, t). && \text{Ass. \ref{ass:ptet}}
\end{align*}

Since there exists at least one $s^1 > t$ such that $P(G^1_{s^1} = 1 | G^2_{g^2} = 1, D^1_t = 0)>0$ as $t < \bar{G}^1_{g^2}$, we have

\begin{align*}  
    & ATT^{dd}_{id}(g^1, g^2, t) \\
    = & ATT^{did}_{id}(g^1, g^2, t) && \text{Eq. \ref{eq:ddid}, } g^1>g^2 \\
    = & ATT_{id}(g^1, g^2, t) - ATT_{id}(g^1, g^2, g^1-1) \\ 
    & - \sum_{s^1 > t}\left(ATT_{id}(s^1, g^2, t) - ATT_{id}(s^1, g^2, g^1-1)\right)P(G^1_{s^1} = 1 | G^2_{g^2} = 1, D^1_t = 0) \\
    = & \sum_{s^1 > t}\big(ATT_{id}(g^1, g^2, t) - ATT_{id}(g^1, g^2, g^1-1) \\ 
    & - ATT_{id}(s^1, g^2, t) - ATT_{id}(s^1, g^2, g^1-1)\big)P(G^1_{s^1} = 1 | G^2_{g^2} = 1, D^1_t = 0) \\
    = & ATT^1(g^1, g^2, t).
\end{align*}

\end{appendices}

\end{document}